\documentclass[10pt]{IEEEtran}

\ifCLASSOPTIONcompsoc
  \usepackage[nocompress]{cite}
\else
  \usepackage{cite}
\fi

\usepackage{balance}  
\usepackage{times}    
\usepackage{amssymb}
\usepackage{url}
\usepackage{wrapfig}
\usepackage{subfigure}
\usepackage{color}
\usepackage{amsmath}
\usepackage{amsfonts}
\usepackage{amsthm}
\usepackage{graphicx}
\usepackage{caption}
\usepackage{verbatim} 
\usepackage{bm}
\usepackage{multirow}
\usepackage{array}
\usepackage{tabu}
\usepackage{booktabs}
\usepackage{cite}
\usepackage{fontenc}
\usepackage[pdftex]{hyperref}
\usepackage[T1]{fontenc}
\usepackage{algorithm}
\usepackage{algpseudocode}

\newfloat{Protocol}{ht}{lop}

\newcommand\abs[1]{\left|#1\right|}

\newcolumntype{L}[1]{>{\raggedright\let\newline\\\arraybackslash\hspace{0pt}}m{#1}}
\newcolumntype{C}[1]{>{\centering\let\newline\\\arraybackslash\hspace{0pt}}m{#1}}
\newcolumntype{R}[1]{>{\raggedleft\let\newline\\\arraybackslash\hspace{0pt}}m{#1}}

\begin{document}

\title{Side-Channel Inference Attacks on Mobile \\Keypads using Smartwatches}

\author{Anindya~Maiti,
	Murtuza~Jadliwala,
	Jibo~He,
	and~Igor~Bilogrevic
	\IEEEcompsocitemizethanks{
	\IEEEcompsocthanksitem A. Maiti, M. Jadliwala and J. He are with the Wichita State University, Wichita, KS 67260 USA \protect\\
	E-mail: a.maiti@ieee.org, murtuza.jadliwala@wichita.edu, jibo.he@wichita.edu
	\IEEEcompsocthanksitem Igor Bilogrevic is with Google, 8002 Zurich, Switzerland \protect\\
	E-mail: ibilogrevic@google.com
	}

}

\IEEEtitleabstractindextext{
\begin{abstract}
Smartwatches enable many novel applications and are fast gaining popularity. However, the presence of a diverse set of on-board sensors provides an additional attack surface to malicious software and services on these devices. In this paper, we investigate the feasibility of key press inference attacks on handheld numeric touchpads by using smartwatch motion sensors as a side-channel. We consider different typing scenarios, and propose multiple attack approaches to exploit the characteristics of the observed wrist movements for inferring individual key presses. Experimental evaluation using commercial off-the-shelf smartwatches and smartphones show that key press inference using smartwatch motion sensors is not only fairly accurate, but also comparable with similar attacks using smartphone motion sensors. Additionally, hand movements captured by a combination of both smartwatch and smartphone motion sensors yields better inference accuracy than either device considered individually.

\end{abstract}

\begin{IEEEkeywords}
Wearables; Smartwatches; Side channel attacks; Keystroke inference.
\end{IEEEkeywords}
}

\maketitle

\section{Introduction} 

The popularity of smartwatches is soaring with more than 45 million devices expected to be shipped by 2017 \cite{canalys}. These devices, typically equipped with state-of-the-art sensors and communication capabilities, will enable a plethora of novel applications, including activity tracking, wellness monitoring and ubiquitous computing. However, the presence of a diverse set of on-board sensors also provides an additional attack surface to malicious applications on these devices. Security and privacy threats on handheld smartphones that take advantage of such sensors as side-channels have received significant attention in the literature. Notable examples include keystroke (or key press) inference \cite{CaiC:2011,OwusuHDPZ:2012,xu2012taplogger}, activity identification \cite{ghosh2014recognizing} and location inference \cite{GaoFSKYL:2014} attacks. As most modern mobile operating systems introduced stringent access controls on front end sensors, such as microphones, cameras and GPS, adversaries shifted attention to sensors which cannot be actively disengaged by users (e.g., accelerometer and gyroscope). Typically, handheld device usage is highly intermittent and such devices spend a majority of time in a constrained (e.g., in users' dress pocket) or activity-less (e.g., on a table) setting where most on-board sensors are partially or completely non-functional, thereby limiting the effectiveness of handhelds in inference attacks. Contrary to this, wearable device usage is much more persistent as they are constantly carried by the users on their body. This makes wearable devices a more desirable platform for a variety of side-channel attacks. If access to wearable sensor data is not appropriately regulated, it can be used as a side-channel to infer sensitive user information.

In this paper, we evaluate the feasibility of side-channel security vulnerabilities in smart wearables by investigating motion-based keystroke inference attacks using smartwatches. More specifically, we evaluate the feasibility and effectiveness of keystroke inference attacks on smartphone numeric touchpads by using smartwatch motion sensors as a side-channel. Numeric touchpads are typically targeted by adversaries for obtaining sensitive information such as security pins and credit card numbers. We propose multiple attacks suitable for three popular typing scenarios. In typing scenarios where key press events can be identified based on surge in motion sensor activity, we use \emph{supervised learning} to infer the key presses. This attack comprises of first training appropriate classification models to learn the uniqueness in wrist motion caused during individual keystrokes depending on known relative location of the key on the screen, and then using the trained classifiers to infer unlabeled (or test) keystrokes. During preliminary experiments, we observed that keystroke induced motion data captured by smartwatch and smartphone sensors differ significantly. Consequently, we thoroughly assess how significantly smartwatch motion sensors elevate the threat of keystroke inference, compared to similar attacks using only smartphone motion data \cite{CaiC:2011,OwusuHDPZ:2012,xu2012taplogger}. We also evaluate the case where the adversary may have gained access to motion sensors on both the smartwatch and smartphone, to see how our attack will perform when motion data from both devices are combined. For the typing scenario where key press events cannot be identified based on the uniqueness of motion sensor activity surge (corresponding to a key press), we present a novel scheme to infer a sequence of key presses based on the \emph{transitional movement} between individual key presses. We evaluate the proposed attacks in both controlled and realistic typing scenarios. We also briefly discuss possible protection measures against such inference attacks that employ motion sensors as side-channels.

The remainder of this article is structured as follows: First we discuss similar side-channel attacks in the literature in Section \ref{related}. Then we give an overview of the threat and adversary model in Section \ref{background}. In Section \ref{sec:expr} we describe the classification-based keystroke inference framework, followed by its evaluation in Section \ref{evaluation1}. In Section \ref{sec:expr2} we describe the relative transitions-based keystroke inference framework, followed by its evaluation in Section \ref{evaluation2}. Finally, we discuss implications, limitations and future research directions.

\section{Related Work}
\label{related}

Inference of private information from various forms of side channels has been an active area of research in the community. Electromagnetic signals emanating from devices have been used to infer private data stored on Smart Cards \cite{Quisquater:2001}, data transmitted on RS-232 cables \cite{Smulders:1990} and content being played on CRT and LED monitors \cite{vanEck:1985,KuhnA:1998}. Recently, Hayashi et al. \cite{hayashi2014threat} showed that it is also possible to remotely reconstruct and eavesdrop on flat panel displays on tablets via measurement of electromagnetic emanations. Similarly, optical emanations from monitors \cite{Kuhn:2002} or from eyes \cite{BackesTDLW:2009} have also be used to infer information such as content being displayed or watched. Acoustic or sound signals emanating from devices such as printers have also been used to infer the content being printed on certain models of dot-matrix printers \cite{BackesDGPS:2010}.

Availability of several high-precision sensors on modern mobile cyber-physical systems such as smartphones have given rise to additional side-channels \cite{uluagac2014sensory}, thus increasing the risk of private information leakage through such side-channels. Past research efforts have shown how malicious applications can misuse their access rights to these sensors in order to execute various imperceptible side-channel attacks by stealthily capturing information from the physical environment. For example, smartphone cameras can be accessed in an unauthorized fashion to infer sensitive information from user keystrokes \cite{FeltFCHW:2011,simon2013pin}.  Unauthorized microphone recordings of ambient sound \cite{schlegel2011soundcomber} provide a rich source of information that can be used to infer sensitive information about a person's daily life. Activities and locations can be inferred based on characteristic ambient sound patterns, e.g. walking on the streets, or eating in a restaurant \cite{rossi2013ambientsense}. Unauthorized access to GPS sensors can pose obvious risks related to loss of location privacy, such as revealing home/work locations, stalking and location-targeted advertisements \cite{iqbal2010privacy}. Advanced learning-based techniques were also proposed for predicting users' future movements from previous tracking records of their location activities \cite{NguyenCWBZ:2012,tiwari2013route,fukano2013next,dewri2013inferring}. Security and privacy risks associated with front-end sensors, such as, microphones, cameras and GPS, have been comprehensively studied because of the hazards apparent to users. 

However, security risks due to sensors obscured from users (e.g., accelerometer, gyroscope, and magnetometer) have largely been overlooked until recently. After modern mobile operating systems introduced user-managed access control on front-end sensors, adversaries shifted attention to sensors which cannot be disengaged by users. It has been shown that malicious applications can track users' movements \cite{HanONPZ:2012,hemminki2013accelerometer} and activities \cite{matic2012speech,zhang2010activity,kwapisz2011activity} by using only smartphone accelerometer readings. It has also been shown that, with the help of standard signal processing and machine learning techniques, it is possible to 
recognize speakers and parse speech by using gyroscopes on modern mobile devices to measure acoustic signals \cite{MichalevskyDG:2014}.

Keystroke inference attacks using side-channel information have received significant attention due to their potentially dangerous consequences. Electromagnetic emanations from external keyboards (both wired and wireless) have been used in the past to infer user keystrokes \cite{Vuagnoux:2009}. However, the requirement of extensive setup and expensive monitoring hardware prevents less sophisticated adversaries from carrying out such attacks. Keystroke inference attacks using audio or acoustic signals \cite{AsonovA:2004,BergerWY:2006,ZhuangZT:2009,HaleviS:2012,ZhuMZL:2014}, on the other hand, have also received significant attention in the literature. Such attacks have proven to be very successful and can be carried out using modest off-the-shelf hardware (e.g., any microphone equipped device). Due to the ubiquitous nature of modern smartphones that are equipped with high-precision microphones, such attacks are much more practical than previously argued. 

However, as touchscreen key press events emanate very weak acoustic signals, inference attacks using them is very difficult. Additionally, requirement of undisturbed eavesdropping is another major obstacle in using electromagnetic and acoustic emanations for such attacks. As a workaround to the above limitations, smartphone motion sensors have been used to recover keystroke events on the device. For instance, \textit{TouchLogger} \cite{CaiC:2011} and \textit{TapPrints} \cite{miluzzo2012tapprints} utilize change in orientation angles of the smartphone, as captured by its accelerometer, to extract appropriate features for keystroke inference. Similarly, \textit{ACCessory} \cite{OwusuHDPZ:2012} also attempts to infer keystrokes using the smartphone accelerometer data by employing multiple supervised learning techniques. Alternatively, \textit{TapLogger} \cite{xu2012taplogger} automates the training and logging phases and attempts to work stealthily on the smartphone. Smartphone motion sensors have also been used to detect keystroke events on other external devices/keyboards in proximity \cite{MarquardtVCT:2012}. Despite these research efforts, side-channel privacy threats (especially, keystroke inference attacks) posed by wearable devices such as smartwatches have received far less attention. Recently, Wang et al. \cite{wang2015mole}, Liu et al. \cite{liu2015good}, Wang et al. \cite{wang2016friend} and Maiti et al. \cite{maiti2016smartwatch} have explored new attacks to infer user keystrokes or key presses on external physical keyboards/keypads by using smartwatch motion sensors. In contrast to these research efforts, our work focuses on keystroke inference on hand-held mobile keypads by employing smartwatch motion sensors.

In our preliminary work in this direction \cite{Maiti:2015:WYT:2802083.2808397}, we designed and evaluated a basic supervised learning-based classification framework to infer keystrokes on a mobile/smartphone keypad. The evaluation was limited to the typing scenarios where key press events were identifiable based on surge in wrist motion activity. In this paper, we present new and improved attack frameworks, which include attack on an additional typing scenario. We made significant changes to the classification-based attack framework, such as enriching the feature vectors and using a robust ensemble classification scheme. Our new relative transition-based attack framework is designed to infer sequences of key presses, where individual key press events cannot be detected by using the unique surge in the wrist motion activity (captured on the smartwatch motion sensor). We conduct extensive empirical evaluation for both frameworks with the help of multi-sensor typing data contributed by human subject participants using two different smartwatch hardware. We also evaluate the efficacy of our attack framework in a non-controlled natural typing setting, Further, we also analyze our attack framework for gain in inference accuracy (if any) when typing-related motion data from both the smartwatch and smartphone is combined.

\section{Attack Description}
\label{background}

\begin{figure}[t]
\vspace{-0.1in}
\centering
	\subfigure[]{
		\includegraphics[height= 1.25in]{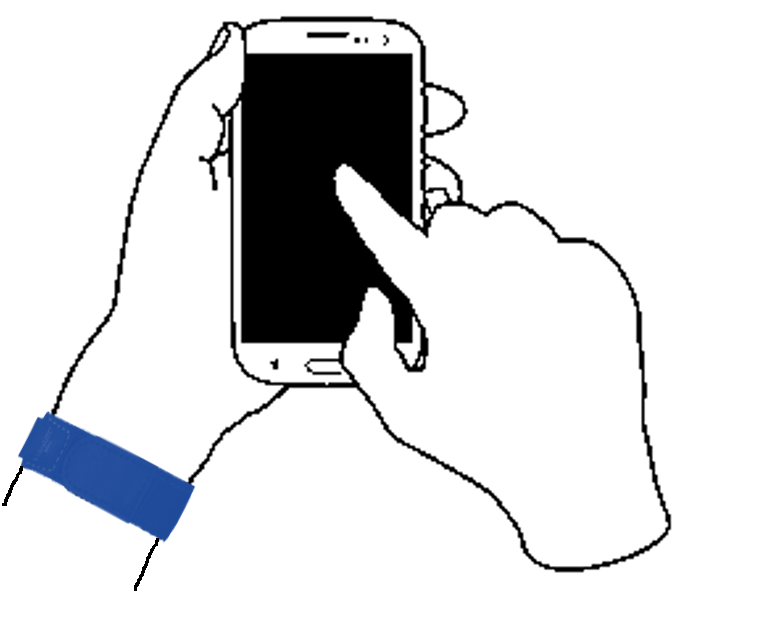}
		\label{fig:same-non-holding}
		}
	\hfill
	\subfigure[]{
		\includegraphics[height= 1.25in]{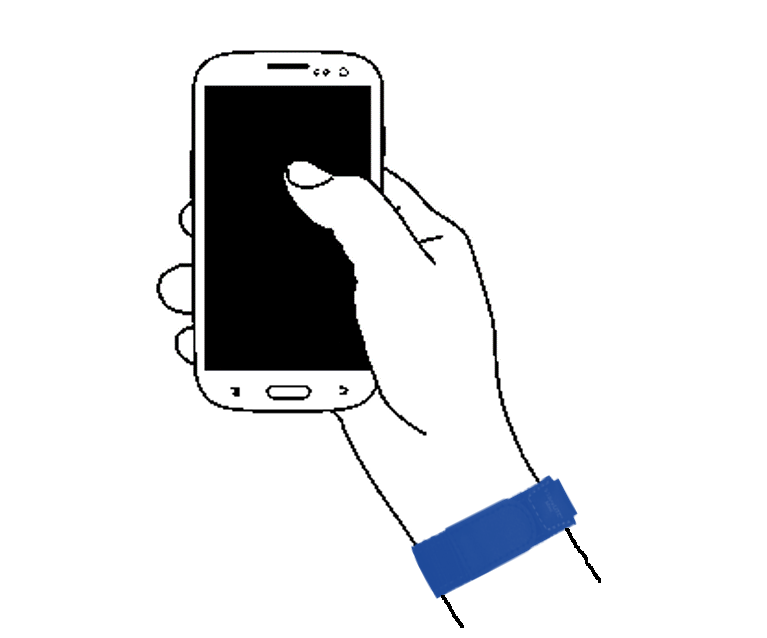}
		\label{fig:same-holding}
		}
	\subfigure[]{
		\includegraphics[height= 1.25in]{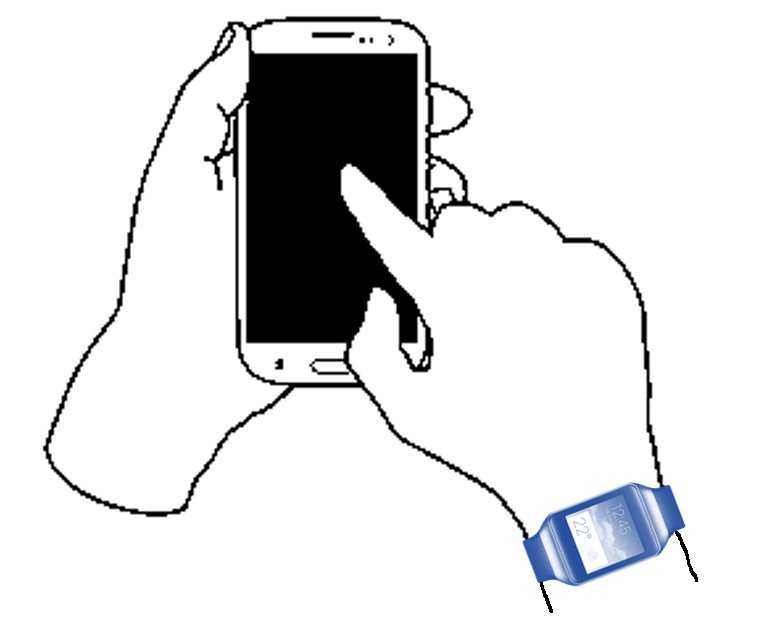}
		\label{fig:diff-non-holding}
		}
	\hfill
	\subfigure[]{
		\includegraphics[height= 1.25in]{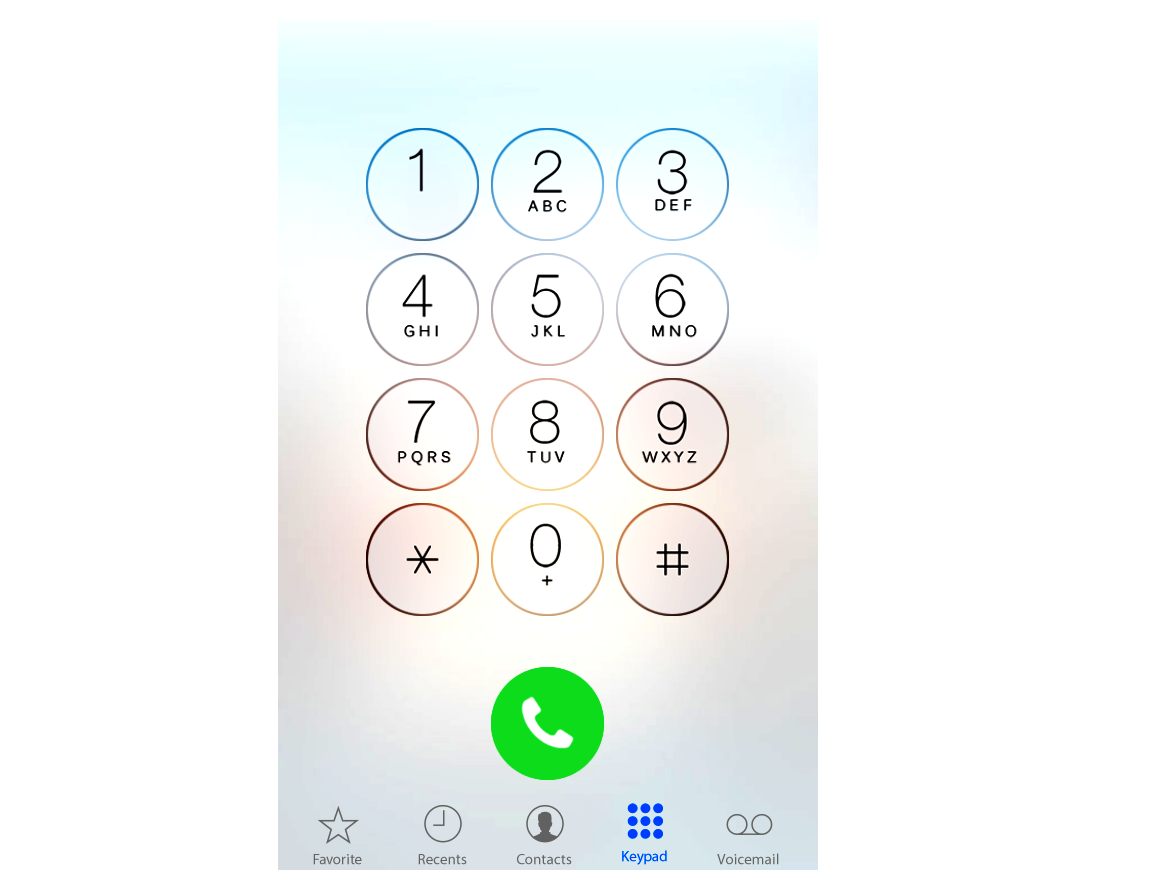}
		\label{fig:keypad}
		}
\vspace{-0.15in}
\caption{Smartwatch and smartphone on (a) \textbf{S}ame \textbf{H}and and \textbf{N}on-\textbf{H}olding \textbf{H}and \textbf{T}yping (SH-NHHT), (b) \textbf{S}ame \textbf{H}and and \textbf{H}olding \textbf{H}and \textbf{T}yping (SH-HHT), (c) \textbf{D}ifferent \textbf{H}and and \textbf{N}on-\textbf{H}olding \textbf{H}and \textbf{T}yping (DH-NHHT), (d) numeric keypad used in our experiments.}
\vspace{-0.0in}
\end{figure}

In this research effort, we focus on three of the most popular typing (or tapping) scenarios in mobile hand-helds or smartphones\cite{ux}. We consider a user typing on a smartphone's numeric touchscreen keypad while wearing a smartwatch on one of his/her hand. In the first case, smartwatch and smartphone are on the same hand and the user types with the hand not holding the smartphone (see Figure \ref{fig:same-non-holding}), also referred by us as \emph{SH-NHHT scenario}. In the second case, smartwatch and smartphone are again on the same hand and the user types with a finger (generally, thumb) of the smartphone holding hand (see Figure \ref{fig:same-holding}), also referred by us as \emph{SH-HHT scenario}. In the above two scenarios, the action of tapping a key on the smartphone keypad results in a unique motion of the wrist (on the smartphone holding hand) for each keystroke, which can be captured by the motion sensors (e.g., accelerometer and gyroscope) of the smartwatch and used to identify the tapped keystroke. 
In the third case, smartwatch and smartphone are on different hands and the user types with the hand not holding the smartphone (see Figure \ref{fig:diff-non-holding}), also referred by us as \emph{DH-NHHT scenario}. Unlike the previous two scenarios, each key tap does not produce a unique motion signature on the wrist of the typing hand (where the smartwatch is situated), and thus it cannot be used to infer the exact keystroke in a fashion similar to the previous two cases.
However, assuming that the relative position of keys on the keypad as per the standard layout shown in Figure \ref{fig:keypad} is known and remains static, we can use the relative transitional movement between taps to infer a (sub)sequence of the tapped keys.

In addition to the above three, other typing scenarios are also possible, for example, typing with both hands and holding (the phone) and typing with non watch wearing hand. However, in order to limit the scope of our study and to clearly demonstrate the keystroke inference threat posed by smartwatches, we only consider the above three typing scenarios (i.e., SH-NHHT, SH-HHT and DH-NHHT) in this work, which also happen to be very widely adopted by smartphone users. We further justify our focus on these three typing scenarios by more precisely determining the percentage of users that employ these typing methods, and as a result, are impacted by the proposed inference attacks. Based on the data available from a study concerning users' smartphone holding and usage behaviors \cite{ux}, 32.83\% of all users hold and use the phone as shown in Figure \ref{fig:same-holding}, 28.44\% of users hold and use the phone as shown in Figure \ref{fig:same-non-holding}, while only 2.11\% of users hold and use the phone as shown in Figure \ref{fig:diff-non-holding}. In this work, we also investigate two variations of scenarios in Figures \ref{fig:same-non-holding} and \ref{fig:same-holding}, where the phone is held in the other hand as shown in Figures \ref{fig:same-non-holding-mirror} and \ref{fig:same-holding-mirror} respectively. Based on \cite{ux}, 16.17\% of all users hold and use the phone as shown in Figure \ref{fig:same-holding-mirror}, while only 7.56\% of users hold and use the phone as shown in Figure \ref{fig:same-non-holding-mirror}. It should be noted that \cite{ux} does not provide any data on which hand (left or right) the smartwatch is traditionally worn, so the above only represents percentages of users based on the hand in which the phone is held and the hand used to type or tap. Now even if the phone holding scenario of Figure \ref{fig:diff-non-holding} is ignored (due to its low usage percentage), our framework can potentially impact at least 44.61\% of users who wear the smartwatch on the left hand, i.e., those who type as shown in Figures \ref{fig:same-non-holding} and \ref{fig:same-holding-mirror}. Similarly, at least 40.39\% of users wearing the smartwatch on the right hand are impacted by our keystroke inference framework, i.e., those who type as shown in Figures \ref{fig:same-holding} and \ref{fig:same-non-holding-mirror}). Moreover, we also show in Section \ref{subsec:variations} that the performance of our framework does not vary significantly between a particular typing scenario and its variation (say, between Figure \ref{fig:same-holding} and \ref{fig:same-holding-mirror}). This shows that a 
significant percentage of users have the potential of being impacted by the proposed keystroke inference framework.

\textbf{Threat Model:} We assume an adversary whose goal is to infer a target's keystrokes on a generic smartphone numeric keypad (as shown in Figure \ref{fig:keypad}), based on the wrist movements perceptible by the target's smartwatch motion sensors. The adversary may gain access to the target's smartwatch by installing a malicious application on it which records the activity of the on-board accelerometer and gyroscope sensors. This step can be achieved by exploiting known software vulnerabilities or by tricking the victim into installing malicious code, e.g., using a trojan software. Based on the fact that most common smartwatch operating systems (e.g., Google's Android Wear, Apple's watchOS, etc.) do not implement access control and/or user notification for motion sensor usage, the malicious application may have unrestricted and undetected access to the on-board accelerometer and gyroscope. As a result, the compromised smartwatch can act as an eavesdropping device which the targets' themselves may place on their wrist, and unsuspectingly have it on their wrist while typing on a smartphone. The malicious application may also maintain a covert communication channel \cite{7444898} with the adversary, and periodically upload the collected wrist motion data on some adversarial server by means of this channel. The use of a covert channel by the trojan is optional. However, if a covert channel is not used, there is a possibility of this information transfer being easily detected and the trojan being inactivated. We assume that the adversary also has sufficient off-site storage and computational resources to download the raw sensor data, extract significant keystroke events, and execute standard machine learning algorithms in order to classify the keystrokes.
For comparison with attacks based on smartphone data, we assume similar adversarial capabilities and actions for the smartphone.

\section{Classification-Based Attack Framework}
\label{sec:expr}

\begin{figure*}[]
\centering
\includegraphics[width=7.0in]{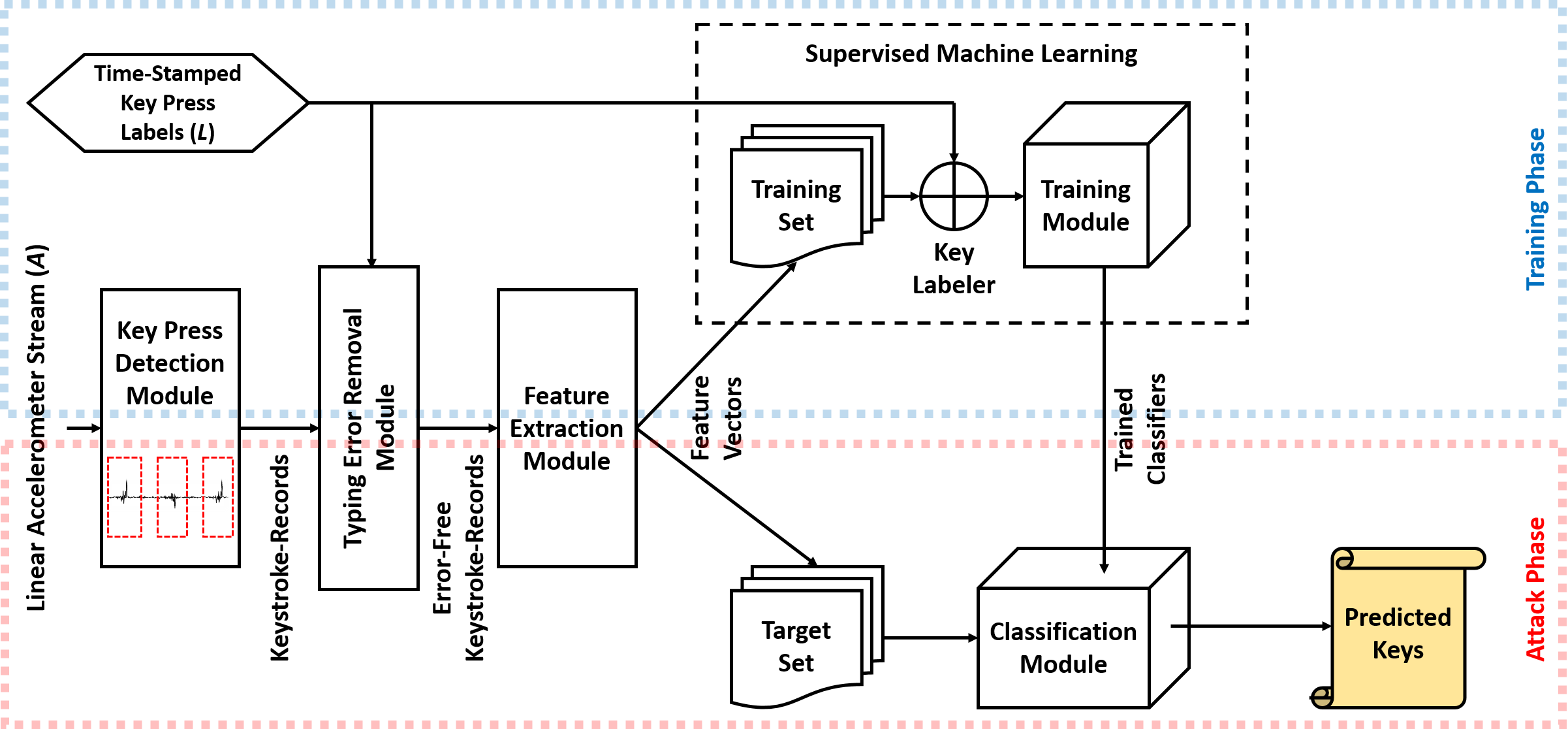}
\caption{\textbf{Overview of the classification-based attack framework for SH-NHHT and SH-HHT typing scenarios.}}
\label{fig:usingclassifier}
\end{figure*}

The linear accelerometer motion sensor found on smartwatches measures the three dimensional acceleration experienced  by the device, excluding the omnipresent force of gravity. During preliminary experimentation with SH-NHHT and SH-HHT scenarios, we observed that key press events can be accurately detected using the surge in linear acceleration during a key press.
Based on the observation that taps on different locations of the smartphone screen produces characteristically unique motions on the wrist, our attack framework leverages on \emph{supervised machine learning} to directly classify the detected key presses. The attack framework (Figure \ref{fig:usingclassifier}) consists of a \emph{learning phase} followed by an \emph{attack phase}. Both phases go through similar steps of \emph{data collection} followed by \emph{feature extraction}, with the learning phase culminating in training (the classifiers), while the attack phase in classification (using the trained classifiers from the training phase). Next, we describe in detail each component of the proposed framework.

\textit{\textbf{Data Collection and Pre-Processing:}}
We developed an application for Android Wear that continuously samples linear accelerometer measurements on the smartwatch, and runs in the background during experiments. Details of the data collection experiments and technical specifications of the hardware used for the experiments are outlined later in Section \ref{ref:exprsetup}. The smartwatch data collection application communicates the linear accelerometer measurements to the host smartphone (with which the watch is paired) using Andorid Wear's Wearable Data Layer API. On the smartphone, another Android application displays a keypad to the users to type sequences of numbers. In the background, the smartphone application chronologically logs all accelerometer measurements received from the smartwatch and any key press events registered on the displayed keypad. It logs two data-streams: (i) timestamped readings of the smartwatch's linear accelerometer ($A$); and (ii) timestamped key press labels ($L$). Both $A$ and $L$ are stored locally on the smartphone (which is also paired with the smartwatch) during the data collection process and retrieved later for offline evaluation. Note that in the attack phase of our experiments we use $L$ only to verify the classification accuracy.

\begin{figure}[]
\centering
\includegraphics[width= 3.4in]{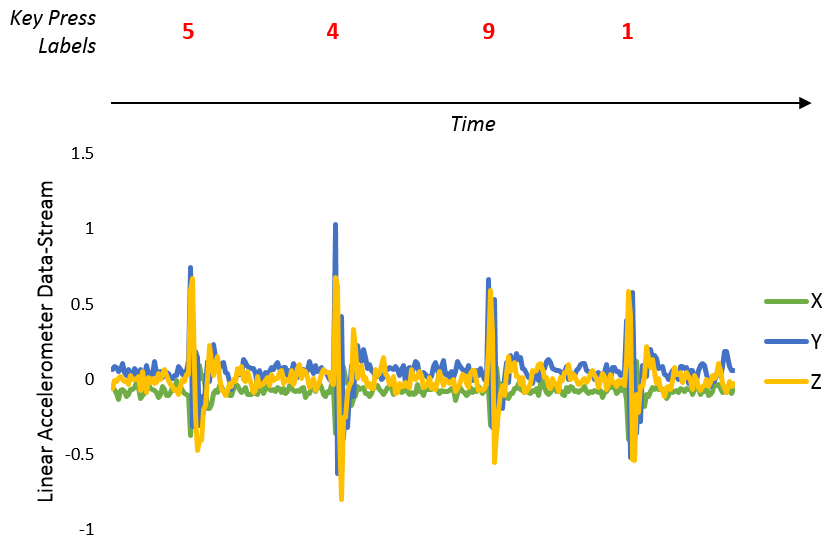}
\caption{\textbf{Time series of key press events in SH-NHHT, and their corresponding effect on linear accelerometer samples.}}
\vspace{-0.0in}
\label{keystrokesfig}
\end{figure}

Due to the absence of labeled data, the attack phase requires an additional key press event detection mechanism. Figure \ref{keystrokesfig} shows a portion of a raw linear accelerometer data-stream, spanning four key press events in the SH-NHHT scenario. As evident from the graph, each tap agitates the linear accelerometer sensor readings on the three axis, with more prominence along the Y-axis and Z-axis than X-axis. SH-HHT data also exhibits similar traits. We apply this observation to model an algorithm for automating the process of key press event detection. Algorithm \ref{keydetectionalgo} sequentially examines the ``energy'' of each sample $i$ in $A$ as the sum of acceleration on the three axis (Equation \ref{eqenergy}). The energy value calculated in Algorithm \ref{keydetectionalgo} is then used in Algorithm \ref{thresholddet} to determine keystroke events.

\begin{equation} \label{eqenergy}
Energy[i] = \abs{\abs{A[i][X]} + \abs{A[i][Y]} + \abs{A[i][Z]}}
\end{equation}

\begin{algorithm}[b]
\caption{Key Press Detection Algorithm}\label{keydetectionalgo}
\begin{algorithmic}[]
\Function{KeyPress\_Detection}{$A^{Target}$}
\State $KeyPresses$ = $\{\emptyset\}$
\State $Threshold$ =  $Set\_Threshold()$
\For {$i=1$ to $N$ ($N$ samples in $A$)}
\If {$Energy[i]$ $\geq$ $Threshold$}
\State $ThisKeyPress$ = $A[i-3]$ to $A[i+15]$
\State Insert $ThisKeyPress$ into $KeyPresses$
\State $i$ = $i + 15$
\EndIf
\EndFor
\EndFunction
\end{algorithmic}
\end{algorithm}

Algorithm \ref{thresholddet} establishes the threshold value as the average peak energy values observed in the time-stamped training set. In the attack phase, once the energy level surpasses the empirically learned threshold (from Algorithm \ref{thresholddet}), a key press event is recognized and a ``keystroke-record'' is saved.  Each keystroke-record is intended to represent the wrist motion pattern of a key press event, and consists of few linear accelerometer readings immediately before and after the key press event is recognized. We empirically observed that the movement due to a key press subsides after approximately 350 $msecs$. Thus, a keystroke-record of eighteen samples (at 50 $Hz$ sampling frequency) sufficiently captures all motion features related to a keystroke. Taking into consideration some of the milder initial motion, we form each keystroke-record as follows: the sample to surpass the energy threshold is preceded by three samples and followed by fourteen samples, chronologically from $A$. After a key press event is recognized and the corresponding keystroke-record is saved, the key press detection algorithm resumes its search for next key press. As multiple samples during a key press may cross the energy threshold, ignoring the fourteen samples following the last keystroke-record ensures that the same key press is not recorded multiple times.

\begin{algorithm}
\caption{Determining Energy Threshold}\label{thresholddet}
\begin{algorithmic}[]
\Function{Set\_Threshold}{$A^{Training}$, $L^{Training}$}
\State $Threshold$ = $0$
\State $KeyPressTime$ =  $0$
\For {$j=1$ to $M$ ($M$ key presses in $L$)}
\State $KeyPressTime$ = $L[j][time]$
\State $ThisKeyEnergy$ = \State $\{Energy[KeyPressTime-1]$ + \State $Energy[KeyPressTime]$ + \State$Energy[KeyPressTime+1]$ + \State $Energy[KeyPressTime+2]\}/4$
\State $Threshold$ = $Threshold$ + $ThisKeyEnergy$
\EndFor
\State $Threshold$ = $Threshold/M$\\
\Return $M$
\EndFunction
\end{algorithmic}
\end{algorithm}

\begin{figure}[b]
\centering
\includegraphics[width= 3.4in]{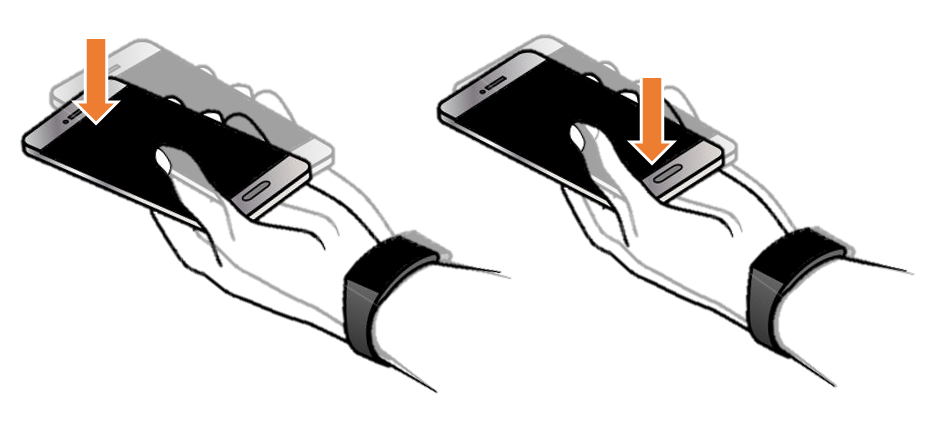}
\caption{\textbf{The intuition behind our classification-based attack is that taps on different locations of the smartphone screen produces characteristically unique motions on the wrist. Accordingly, taps on each number on the keypad should be identifiable based on the uniqueness in the resultant wrist motion.}}
\vspace{-0.0in}
\label{intuition1}
\end{figure}

\textit{\textbf{Feature Extraction:}}
Our proposed attack infers the numeric key that was pressed based on features of the underlying physical event of wrist motion caused during typing (or tapping) on a smartphone. The features of a keystroke-record must be able to capture as many attributes as possible about the underlying three-dimensional movement caused by a key press. A properly designed feature vector should be similar with other feature vectors of the same key, simultaneously being distinguishable between feature vectors of other keys. We observed that, based on the location of a key on the screen, the degree of movement caused by a tap varies on each of the $X$, $Y$, and $Z$ axis of the linear accelerometer (Figure \ref{intuition1}). Interestingly, this movement remains fairly consistent for the same key.

In our preliminarily work \cite{Maiti:2015:WYT:2802083.2808397}, we used only 54 basic time domain features of the accelerometer data to identify the uniqueness of each key (and the corresponding key press event), and found those features to be reasonably useful for keystroke inference. In this work, we expand that to a more comprehensive set of features, employing both time and frequency domain features, with a total of 155 different features in our feature vector for each key. We continue to use time domain features of individual axis such as minimum and maximum magnitudes, squared sum of magnitude data below 33 percent and above 67 percent of maximum magnitude (to measure the duration of major and minor movements), position of maximum and minimum magnitude samples, mean, median, variance, standard deviation, skewness (measure of any asymmetry) and kurtosis (to measure any peakedness), raw accelerometer readings, and their first order numerical derivatives (to measure the rate of change of energy). We also use inter-axis time domain features to capture the correlation between movement on the three axis, such as minimum and maximum magnitudes across all three axis, Frobenius norm, Infinity norm, 1-norm, Euclidean norm, and axis with highest and lowest magnitude for each time sample. Along with the time domain features, we also capture frequency domain features by computing the Fast Fourier Transform (FFT) of individual axis readings of the keystroke-record. The frequency domain features are necessary to identify the different rebounding (or oscillatory) motion of the wrist. Note that in the learning phase, the feature vectors are also labeled, using the timestamped key press labels ($L$) recorded by the data collection application.

\textit{\textbf{Training and Classification:}}
We model the keystroke inference problem as a \emph{multi-class classification} problem. Labeled feature vectors are used to train classifiers in the \emph{learning phase}, whereas unlabeled feature vectors are mapped to the ``closest'' matching class by the already trained classifiers during the \emph{attack or test phase}. To train our classifiers, we initially tested five different classification algorithms that are appropriate given the properties of our features: (i) \emph{simple linear regression (SLR)}, (ii) \emph{random forest (RF)}, (iii) \emph{k-nearest neighbors (k-NN)}, (iv) \emph{support vector machine (SVM)}, and \emph{bagged decision trees (BDT)}. However, each of these classification techniques has its own advantages and shortcomings, leading us to adopt an \emph{ensemble classification approach}. Ensemble approaches have proven to be more accurate and robust than any single classification algorithm \cite{caruana2004ensemble,jahrer2010combining}. We consider an extremely broad set of classification algorithms in our ensemble method, as a result of which, the errors made by constituting classifiers are highly uncorrelated. We include parametric algorithms (SLR, SVM), as well as non-parametric algorithms (k-NN, RF, BDT). Our ensemble method involves both linear (SLR, SVM) and non-linear (k-NN, RF, BDT) techniques. Moreover, RF and BDT are strong ensemble classifiers in themselves which makes our classification framework even more robust. Such a diverse set of classification algorithms increases the likelihood of improvements in classification accuracy over a single algorithm.

\begin{figure}[t]
\centering
\includegraphics[width= 3.4in]{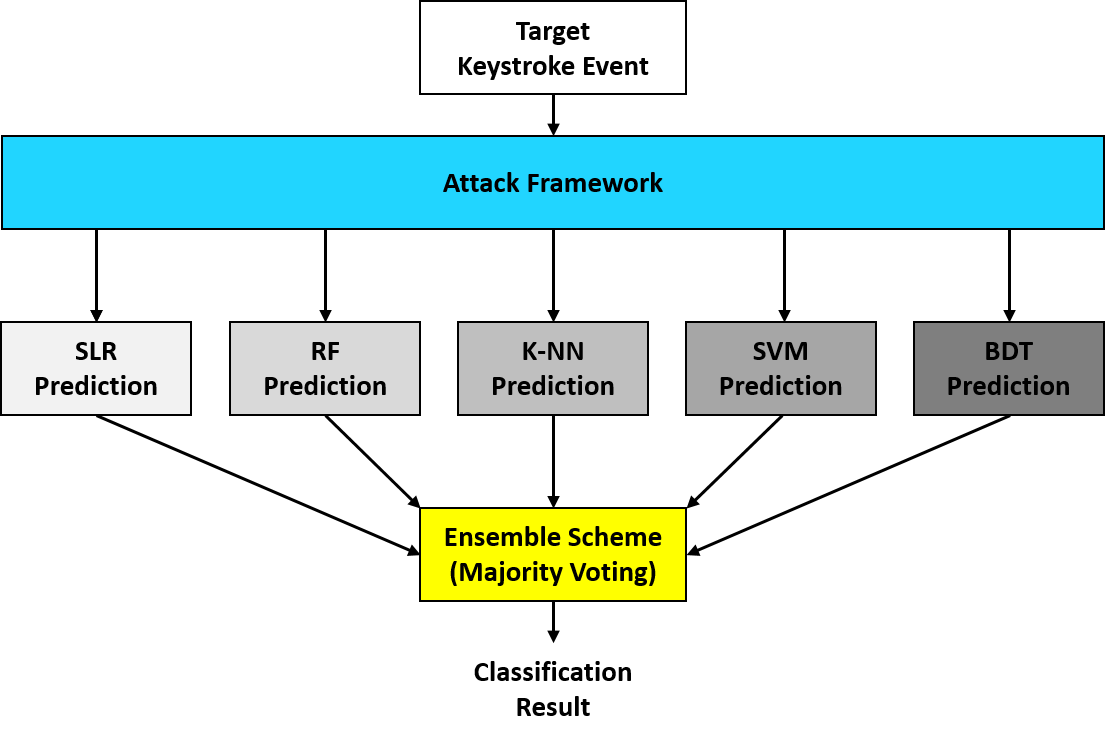}
\caption{\textbf{Ensemble classification scheme used in the attack phase is robust and generally more accurate than a single classification algorithm.}}
\vspace{-0.0in}
\label{ensemble}
\end{figure}

During the training phase, multi-class classifiers of each constituting classification algorithm are trained separately using the labeled training data. After all the classifiers have been trained using the labeled data, feature vectors of unlabeled keystroke-records are classified using these trained classifiers (in the attack phase) using an ensemble strategy. Finally, a \emph{majority wins} ensemble strategy is used to determine the final classification result (Figure \ref{ensemble}).

\section{Evaluation of Classification-Based Attacks}
\label{evaluation1}

In this section, we present the findings from our evaluation of the classification-based attack framework.

\subsection{Experimental Setup}
\label{ref:exprsetup}
Our initial data collection experiments involve 12 participants, aged between 19-32 years. The identity of these participants are anonymized as $P_1$, $P_2$, \ldots, $P_{12}$. We employ a Samsung Gear Live smartwatch equipped with an InvenSense MP92M 9-axis Gyro + Accelerometer + Compass sensor. Smartwatch was worn on left hand for SH-NHHT (Figure \ref{fig:same-non-holding}) and on right hand for SH-HHT (Figure \ref{fig:same-holding}). Participants use the virtual numeric keypad of a Motorola XT1028 smartphone (Figure \ref{fig:keypad}) for typing. Linear accelerometer of the smartwatch was sampled at 50 Hz. We used the Weka 3.7.12 \cite{weka} libraries for both training and testing the classifiers. MATLAB R2014a was used to compute most of the time and frequency domain features. 

We also evaluate the performance of our keystroke inference framework in several additional settings: (i) a more natural or uncontrolled typing scenario (Section \ref{realisticsetting}), (ii) using a different smartwatch hardware (Section \ref{crossdevice}), (iii) employing an additional type of motion sensor, i.e., gyroscope (Section \ref{supplementary}), and (iv) typing on a QWERTY or alphabetic keypad (Section \ref{qwerty}). For these last four experimental settings, we collected additional data from different sets of participants, and in certain cases using a different smartwatch and/or smartphone hardware. The participant and data collection procedure details for these additional experimental settings appear in their respective sections.

\subsection{Constructing and Testing the Classifiers}
\label{ctc}
We construct our classifiers based on different training datasets of labeled keystroke-records generated by the participants. An audio stream of uniformly distributed random numbers between 0 to 9 guided the participants in typing. To prevent fatigue, participants were given optional breaks, during which they were allowed to set down the phone on the table and some participants even went out of the room. However, they returned to approximately the same holding position after the break. We comparatively evaluate the classification accuracy (the percentage of correct prediction divided by the total number of predictions) of our classifiers for the following three training/testing scenarios:
\begin{itemize}
\setlength\itemsep{0 em}
\item
\textbf{\textit{One vs. One}}: In this case, we measure the percentage of successful inferences on an individual participant, with classifiers trained from the training set of the same participant. Target set size is 100 (10 per key) and training set size is 200 (20 per key). \textit{One vs. One} is not only a best case scenario, but also represents how the attack will perform if the adversary is able to collect target-specific training data.
\item
\textbf{\textit{One vs. Rest}}: In this case, we measure the percentage of successful inferences on an individual participant, with classifiers trained from the training set of the rest of the participants (not including the target participant). Target set size is 100 (10 per key) and training set size is 2200 (220 per key). \textit{One vs. Rest} is a typical scenario where the adversary has a target, but is unable to obtain labeled training data from the target.
\item
\textbf{\textit{All vs. All}}: In this case, we measure the percentage of successful inferences on all participants, with classifiers trained from training set of all participants. Target set size is 1200 (120 per key) and training set size is 2400 (240 per key). \textit{All vs. All} is helpful in understanding how our attack framework will perform if the adversary constructs a heterogeneous training data set to infer keystrokes from multiple non-specific targets.
\end{itemize}

\begin{figure}[b]
\centering
\includegraphics[width= 3.49in]{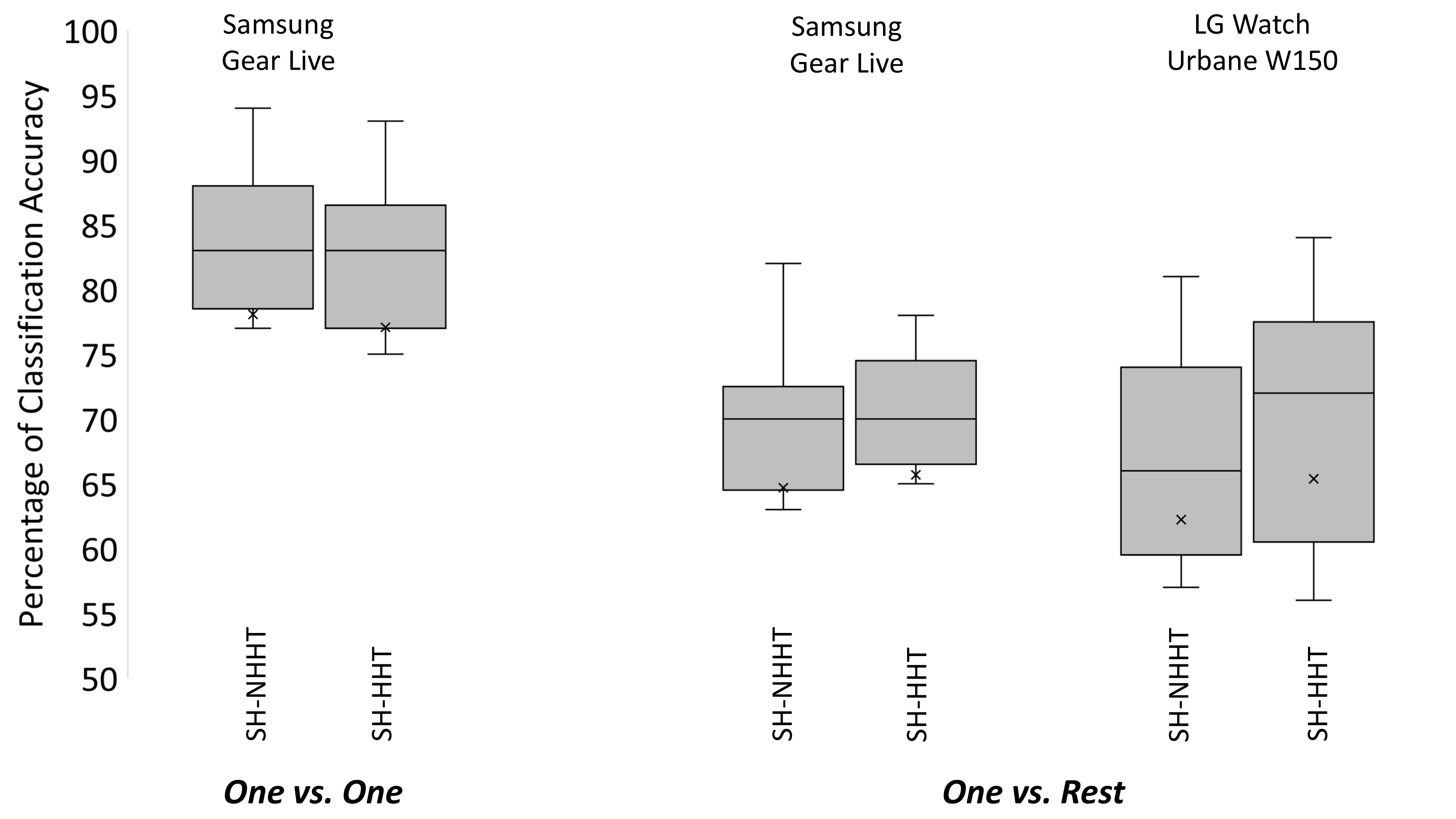}
\caption{\textbf{Classification accuracy for \textit{One vs. One} and \textit{One vs. Rest} using two different smartwatches (Samsung Gear Live and LG Watch Urbane W150).}}
\label{new-2watchclass}
\vspace{-0.0in}
\end{figure}

Classification results for \textit{One vs. One} are shown in Figure \ref{new-2watchclass}. \textit{One vs. One} classification accuracy ranged fairly high between 94\% and 77\% for SH-NHHT, and between 93\% and 75\% for SH-HHT, with an average of 84.58\% and 83.5\%, respectively. However, classification accuracy drops noticeably in \textit{One vs. Rest}. As shown in Figure \ref{new-2watchclass}, \textit{One vs. Rest} classification accuracy ranged between 82\% and 63\% for SH-NHHT, and between 78\% and 65\% for SH-HHT, with an average of 70.08\% and 71.16\%, respectively. The achieved \textit{All vs. All} classification accuracy was 88.16\% and 85.83\% for SH-NHHT and SH-HHT, respectively. Overall, these results validate our claim that smartwatch motion sensors are a feasible side-channel for inferring keystrokes on mobile touchpads.

\begin{table}[t]
  \centering
  \caption{Mean computation time observed in each training/testing scenario. All measurements are in seconds.}
    \begin{tabular}{lrrrrrr}
    \toprule
          & SLR   & RF    & K-NN  & SVM   & BDT   & Total \\
    \midrule
    \textit{One vs. One} & 191   & 234   & 98    & 95    & 65    & 683 \\
    \textit{One vs. Rest} & 166   & 365   & 398   & 184   & 167   & 1280 \\
    \textit{All vs. All} & 504   & 617   & 441   & 271   & 218   & 2051 \\
    \bottomrule
    \end{tabular}%
  \label{comptime}%
\end{table}%

We also recorded the mean computation time  in each training/testing scenario (Table \ref{comptime}). All training and testing operations were executed on a laptop featuring a 2.7 $GHz$ dual-core Intel i5 processor and 8 GB of working memory. Due to the use of ensemble classification technique, the total computation time is the sum of time taken by the five constituting classification algorithms. The average total computation time in \textit{One vs. One} scenario was less then 12 minutes, about 21 minutes in  \textit{One vs. Rest}, and about 34 minutes in \textit{All vs. All} scenario.  Moreover, the total computation time can be further reduced if the different classification algorithms are executed in parallel (with suitable hardware support). These results show that the above keystroke inference attacks can be carried out by an attacker using reasonable computation resources in a fairly short amount of time.

\subsection{Reduced Sampling Frequency:} 
\begin{figure}[]
\centering
\includegraphics[width= 3.45in]{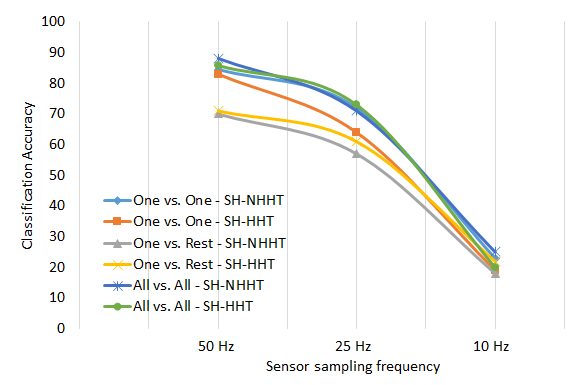}
\caption{\textbf{Classification accuracy dropped when sampling rate was reduced, results averaged over all 12 participants.}}
\label{reducedsampling}
\vspace{-0.0in}
\end{figure}

We also briefly investigate how our attack will perform at reduced sampling rate (25 $Hz$ and 10 $Hz$), a more realistic scenario for low-cost wearables, equipped with less precise sensors. We repeat the experiments outlined in Sections \ref{ref:exprsetup} and \ref{ctc} with smartwatch data sampled at a reduced frequency, and Figure \ref{reducedsampling} shows the drop in accuracy of our attacks for both the SH-NHHT and SH-HHT scenarios. For example, \textit{One vs. One} classification accuracy in SH-NHHT dropped from 84.58\% to 72\% when sampling frequency was reduced to 25 $Hz$ and to 23\% when sampling frequency was reduced to 10 $Hz$. Similarly, the other scenarios also observed drop in classification accuracy with reduction in sampling frequency, but percentage of successful classification can be considered fairly substantial even at a sampling frequency of 25 $Hz$.

\subsection{Comparison with Smartphone-Based Attacks}
\label{supplementary}
\begin{figure}[b]
\centering
\includegraphics[width= 3.45in]{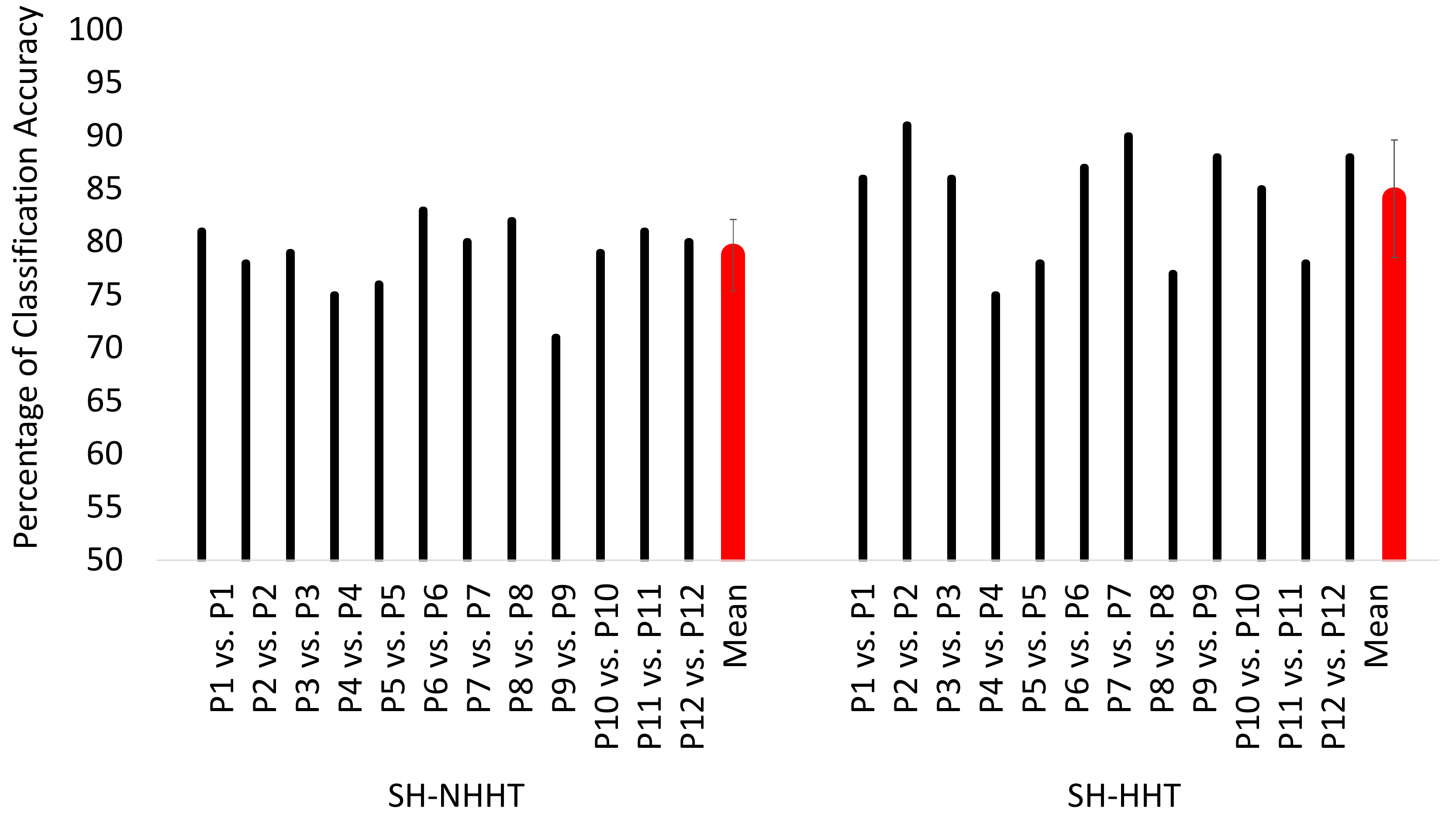}
\caption{\textbf{Classification accuracy for \textit{One vs. One} using smartphone data.}}
\label{ovso-sp}
\vspace{-0.0in}
\end{figure}

\begin{figure}[]
\centering
\includegraphics[width= 3.45in]{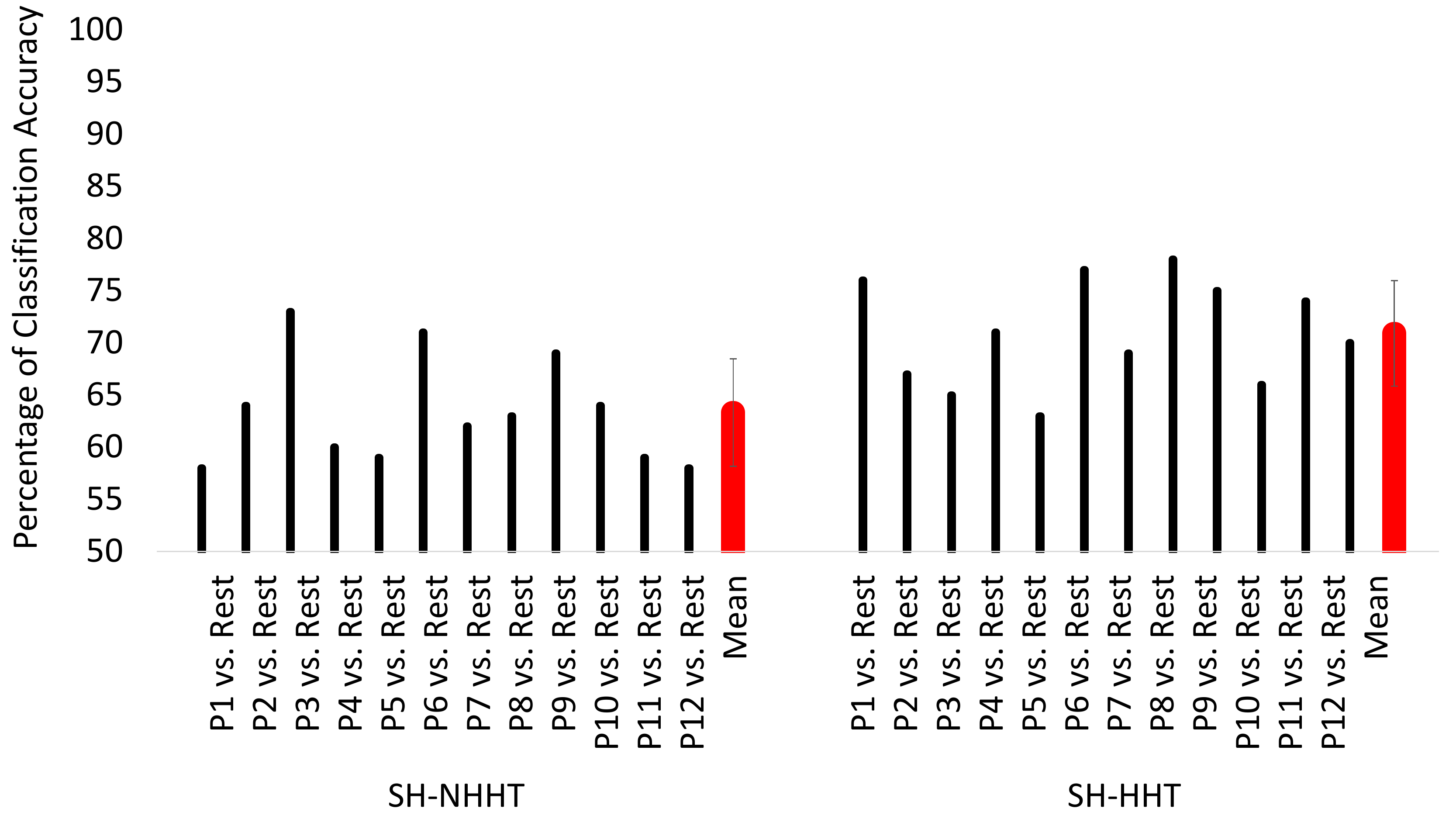}
\caption{\textbf{Classification accuracy for \textit{One vs. Rest} using smartphone data.}}
\label{ovsr-sp}
\vspace{-0.0in}
\end{figure}

\begin{figure}[]
\centering
\includegraphics[height= 1.9in]{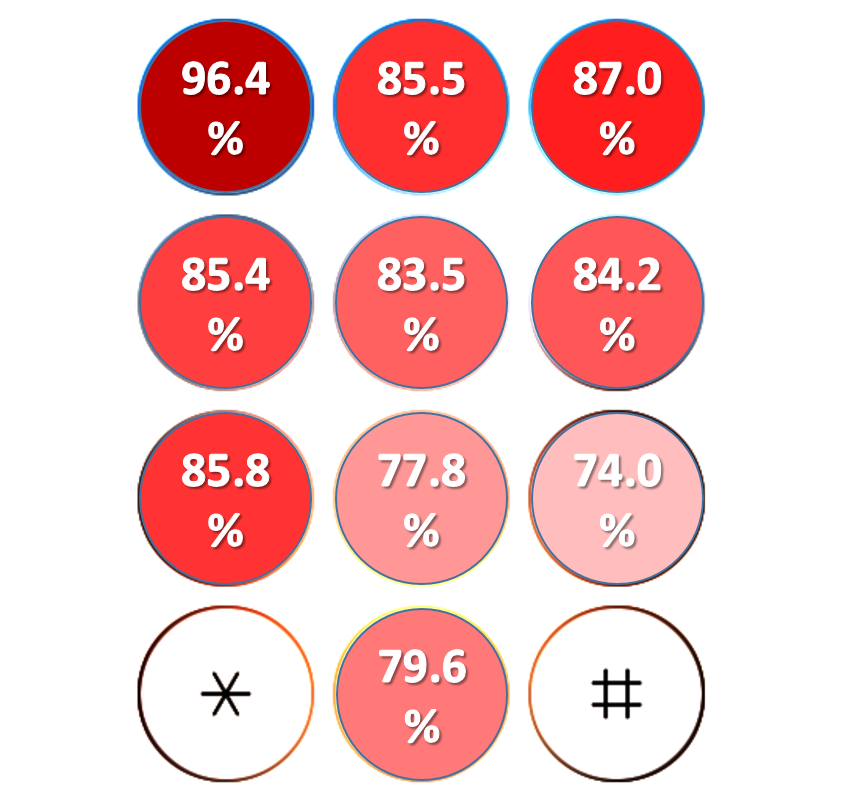}
\caption{\textbf{\textit{All vs. All} classification accuracy for individual keys in SH-HHT using smartphone data, results averaged over all 12 participants.}}
\label{fig:SH-HHTpattern}
\vspace{-0.0in}
\end{figure}

Previous research efforts on keystroke inference attacks by using smartphone sensor data \cite{xu2012taplogger,miluzzo2012tapprints} (or data collected from the target's smartphone sensors) also used similar learning-based multi-class classification frameworks. This motivated us to apply our attack framework on smartphone data and compare the results with those carried out using smartwatch data. This enables us to understand how much more or less vulnerable a motion sensor-based side-channel originating on a smartwatch makes us, as compared to known motion-based side-channels on the target users' smartphone. We conduct similar experiments (as in Sections \ref{ref:exprsetup} and \ref{ctc}) by using smartphone linear accelerometer data sampled at 50 Hz, rather than using the smartwatch data. Figures \ref{ovso-sp} and \ref{ovsr-sp} shows the accuracy of our attack for SH-NHHT and SH-HHT scenarios. 
On comparing with previous results from Section \ref{ctc}, it can be observed that the keystroke inference attacks in SH-NHHT resulted in slightly better average classification accuracy when smartwatch motion data was used. Whereas in SH-HHT, classification accuracy results are mixed, and nearly equal, for both the smartwatch and smartphone data. In summary, these results demonstrate that the threat of motion-based keystroke inference may be increased in certain typing scenarios due to smartwatches.
An interesting pattern of classification accuracy can be observed (see Fig. \ref{fig:SH-HHTpattern}) for inference using only smartphone data in SH-HHT. We observe that the classification accuracy for certain keys (based on their location) are distinctly higher than others. Interestingly, this occurrence is not recognizable for the smartwatch dataset. This may be due to the fact that keys farther away from the thumb impels the user to bend the phone towards the thumb. As a result, significantly greater movement of the phone occurs, compared to keys that are near the thumb.

\begin{figure}[]
\centering
\includegraphics[width= 3.49in]{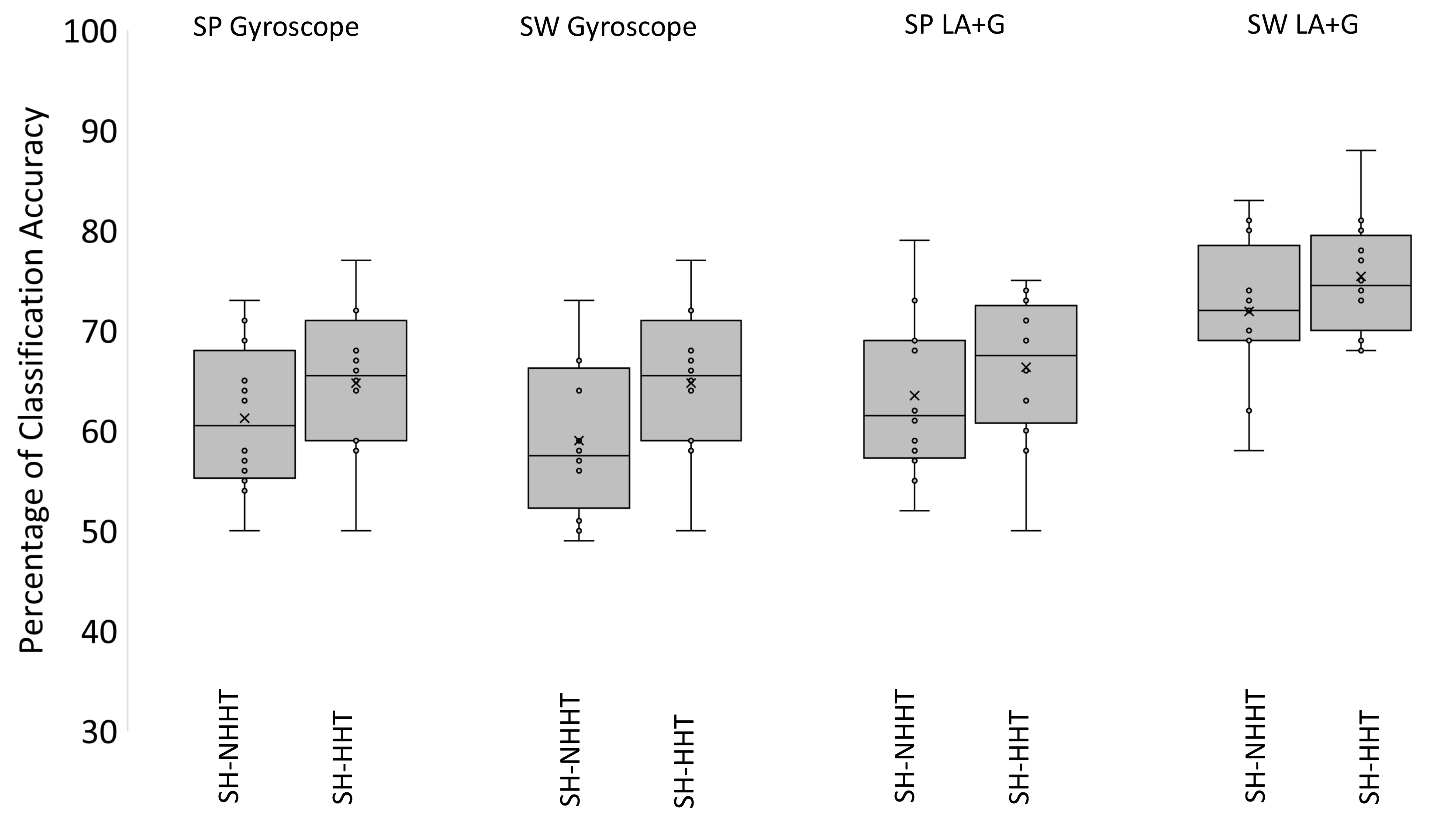}
\caption{\textbf{\textit{One vs. Rest} classification accuracy using only gyroscope features, and in combination with linear accelerometer features. Results compared between smartwatch and smartphone.}}
\label{new-gyro2}
\vspace{-0.0in}
\end{figure}

In order to conduct an exhaustive comparison between the inference threat posed by different motion sensors present on a smartwatch and smartphone, we carry out additional experiments using the gyroscope data which is another widely studied side-channel for keystroke inference \cite{CaiC:2011,miluzzo2012tapprints,cai2012practicality}. Due to the absence of a gyroscope sensor on the Motorola XT1028, we used another smartphone for this experiment, namely a Motorola XT1096 (paired with the Samsung Gear Live). The same experiment as above was repeated by 12 new participants, each typing 100 randomly dictated numbers. For this experiment, we recorded keystroke related motion data, comprising of both linear accelerometer and gyroscope measurements, from both the smartphone and the smartwatch. We derived 59 time and frequency domain features from the three-dimensional gyroscope data of both devices, such as minimum and maximum values, the mean value, variance, skewness, kurtosis, vertex angles, number of spikes, peak intervals, attenuation rate, etc. These features were selected from the literature on activity detection \cite{altun2010human,zhang2011feature} and keystroke inference \cite{cai2012practicality}. Figure \ref{new-gyro2} shows the \textit{One vs. Rest} classification accuracy results when solely the gyroscope features are used compared to when they are used in combination with features derived from the linear accelerometer measurements. The mean classification accuracy is marginally lower when using only the smartwatch gyroscope, compared to the smartphone gyroscope (SH-NHHT: 59.91\% vs. 61.25\%, SH-HHT: 59.66\% vs. 64.75\%). However, we can observe that after combining multiple motion sensors (linear accelerometer and gyroscope) on a device, the keystroke inference threat on the smartwatch is greater than the one on the smartphone (mean classification accuracy, SH-NHHT: 69.91\% vs. 61.41\% and SH-HHT: 71\% vs. 66.08\%).

\subsection{Combining Smartwatch and Smartphone Data}
\label{smartphonecompare}
After comparing the accuracy of keystroke inference attacks using individually both the smartwatch and smartphone motion data, we were intrigued to study the impact of combining or fusing motion sensor data from both devices in order to further reduce the number of classification errors. As most modern smartwatch operating systems and applications require the watch to be paired with a smartphone, such an attack is quite realistic. The feature vectors of \emph{same} keystroke-records from both the devices were merged to obtain new feature vectors containing 310 features. We rebuild the classifiers with the larger feature vectors, and re-ran the previous experiments (as outlined in Sections \ref{ref:exprsetup} and \ref{ctc}). Results of these experiments (outlined in Table \ref{combine1}) show that indeed accuracy improved when the features from both smartwatch and smartphone were combined. For example, the \textit{One vs. One} classification accuracy in SH-NHHT was 90.66\%, compared to 83.5\% and 84.0\% when individual smartwatch or smartphone data were used, respectively. Similar improvements can be observed in other scenarios as well. However, the improvement was relatively marginal, which can be attributed to the convergence in the learning process. Therefore, combining or fusing data from both smartwatch and smartphone may be more beneficial when the adversary has fewer training data.

\begin{table}[]
  \centering
  \caption{\textbf{Classification accuracy after combining features from both smartwatch and smartphone, results averaged over all 12 participants.}}
  \vspace{0.05in}
    \begin{tabu} to \columnwidth {X[3]X[5c]X[5c]}
    \toprule
         & \textbf{SH-NHHT Combined} (Smartwatch Only, Smartphone Only) & \textbf{SH-HHT Combined} (Smartwatch Only, Smartphone Only) \\
    \midrule
    \textit{One vs. One}		& \textbf{88.91\%} (84.5\%, 78.7\%) &	\textbf{90.66\%} (83.5\%, 84.0\%)\\
    \midrule
    \textit{One vs. Rest}		& \textbf{71.59\%} (70.0\%, 63.3\%) &	\textbf{74.29\%} (71.1\%, 70.9\%) \\
    \midrule
    \textit{All vs. All}			& \textbf{88.65\%} (88.1\%, 85.5\%) &	\textbf{89.78\%} (85.8\%, 86.8\%) \\
    \bottomrule
    \end{tabu}
  \label{combine1}
\end{table}

\begin{figure}[t]
\centering
\includegraphics[width= 3.45in]{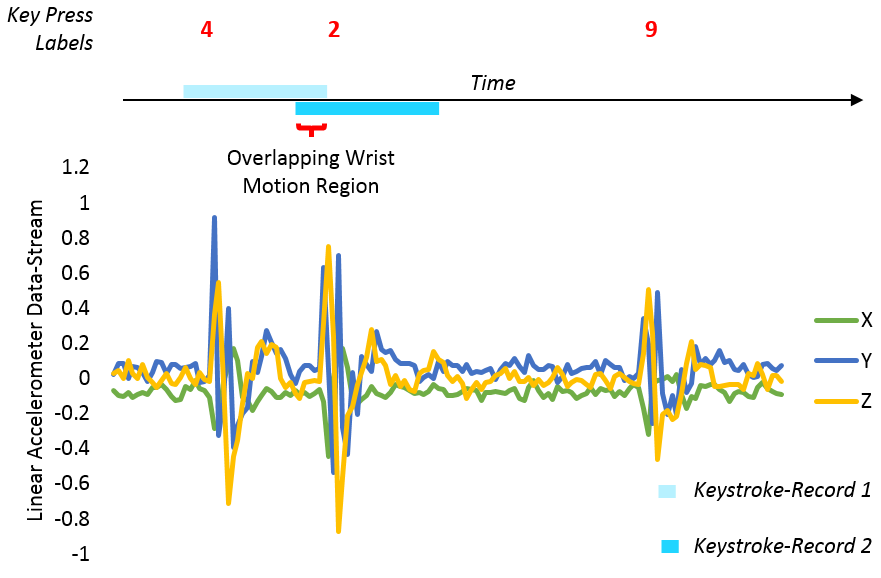}
\caption{\textbf{An example where rebounding motion of a key press overlapped with the next key press.}}
\vspace{-0.0in}
\label{overlap}
\end{figure}

\subsection{A More Realistic Setting: Natural or Non-Controlled \\Typing}
\label{realisticsetting}
In all of the experiments so far, the participants were being directed (to tap) by an audio stream. Because participants have to hear the audio and then act on it, a minor delay or disturbance may be introduced in each key press. Moreover, such a kind of typing or tapping does not invoke (and capture) users' natural typing behavior and speed.  To evaluate a more natural typing behavior, we conduct another experiment where a new set of 10 participants were instructed to type their phone number followed by their residential zip code (a total of 15 numbers). These two pieces of information can be readily recollected by participants, thus eliminating any delay and/or disturbance while typing. This also enables us to capture more realistic typing or tapping data from users. However, for prediction we continue to use the classifiers trained earlier in the guided experiments (Section \ref{ctc}). The new data was processed by the same attack framework to extract keystroke-records and build feature vectors. We obtained a mean classification accuracy of 52\% and 61\% for SH-NHHT and SH-HHT, respectively. It was observed that the primary cause of drop in classification accuracy resulted from faster typing, where the rebounding motion of few key presses overlapped with their next key press (see Figure \ref{overlap}). Such instances were observed more often when two consecutive key presses were for number adjacent to each other on the keypad. Although the classification accuracy of naturally typed numbers is not as high as in the guided experiments, it is high enough to be a significant threat.

\subsection{Cross Device Performance}
\label{crossdevice}
In order to further evaluate how the proposed attack framework performs across different commercial wrist wearable or smartwatch hardware, we test our trained classifiers (from Section \ref{ctc}) on keystroke motion data obtained from a smartwatch of a different make and model. This simulates a situation where an adversary trains classification models using one type of smartwatch hardware and then employs those models to infer the keystrokes of a target user who is using a completely different (possibly, unknown) smartwatch. Such a situation is much more realistic.
For this set of experiments, we used a LG Urbane W150 smartwatch that has a InvenSense M651 accelerometer and gyroscope sensor and collected keystroke motion data from 12 completely new participants. 
Motion data corresponding to 100 keystrokes were collected from each of the new participants, and tested using the classifiers trained earlier in Section \ref{ctc}. The new data was collected at the same sampling frequency of 50 $Hz$. Results (Figure \ref{new-2watchclass}) show that while mean classification accuracy  dropped slightly on the Urbane W150 (SH-NHHT: 70.08\% vs. 67.41\%, SH-HHT: 71.16\% vs. 70.83\%), the variance is significantly lower in case of the Gear Live (SH-NHHT: 26.62 vs. 56.26, SH-HHT: 17.24 vs. 77.0). Although such a trend is intuitive, it nevertheless shows that keystroke inference using the propose framework is still feasible with reasonable accuracy even in such a realistic setting.

\subsection{Extending to QWERTY Keypads}
\label{qwerty}
Up until this point, our primary focus has been keystroke inference attacks on numeric mobile keypads. We now briefly investigate how our proposed attack framework performs against alphanumeric mobile keypads with the standard QWERTY layout. Intuitively, as the keys on a standard smartphone QWERTY keypad are relatively smaller and placed closer to each other (compared to keys on the numeric keypad), keystroke prediction may suffer from high confusion with neighboring keys \cite{miluzzo2012tapprints}. We collected 1248 alphabet keystrokes from a completely new set of 12 participants using the LG Urbane W150 smartwatch, with equal distribution of alphabets (48 each). We then re-ran the training and attack modules in the \textit{One vs. Rest} setting, with 75\% data used for training and 25\% data used for testing. Table \ref{tab:qwerty} summarizes the classification accuracy of the 26 alphabets, along with two most confused keys predicted for each alphabet. As anticipated, the classification accuracy is significantly lower on the QWERTY keypad (compared to the numeric keypad), with an average accuracy of 30.44\%. While the low classification accuracy of individual keys is prohibitive in carrying out effective inference attacks, it is important to note that the most confused keys are usually neighboring to the actual key. It is possible that we may be able to further improve the accuracy of these inference attacks by analyzing keyboard characteristics and/or performing a dictionary-based search \cite{MarquardtVCT:2012,maiti2016smartwatch}. 

\begin{table}[h]
  \centering
  \caption{\textbf{Classification accuracy of the 26 alphabets (in percent), along with two most confused keys predicted for each alphabet. Results averaged over all 12 participants.}}
    \begin{tabu} to \columnwidth {X[2.5c]|X[2c]|X[2c]}
    \toprule
    \textbf{Accuarcy} & \textbf{1st Confusion} & \textbf{2nd Confusion} \\
    \midrule
    a: 41.66 & s: 25.00 & z: 16.66 \\
    b: 25.00 & v: 33.33 & g: 16.66 \\
    c: 33.33 & f: 25.00 & v: 25.00 \\
    d: 16.66 & s: 33.33 & c: 25.00 \\
    e: 33.33 & w: 33.33 & d: 16.66 \\
    f: 16.66 & d: 33.33 & v: 16.66 \\
    g: 25.00 & h: 58.33 & b: 8.33 \\
    h: 33.33 & g: 25.00 & n: 25.00 \\
    i: 33.33 & o: 25.00 & u: 25.00 \\
    j: 16.66 & h: 16.66 & k: 16.66 \\
    k: 16.66 & j: 41.66 & m: 16.66 \\
    l: 25.00 & k: 25.00 & o: 16.66 \\
    m: 33.33 & k: 25.00 & n: 8.33 \\
    n: 25.00 & h: 16.66 & m: 16.66 \\
    o: 33.33 & i: 16.66 & l: 8.33 \\
    p: 50.00 & o: 25.00 & i: 8.33 \\
    q: 41.66 & a: 16.66 & w: 16.66 \\
    r: 25.00 & e: 33.33 & f: 16.66 \\
    s: 16.66 & x: 25.00 & z: 25.00 \\
    t: 33.33 & f: 16.66 & h: 16.66 \\
    u: 41.66 & h: 33.33 & k: 16.66 \\
    v: 25.00 & c: 33.33 & b: 25.00 \\
    w: 41.66 & q: 25.00 & e: 8.33 \\
    x: 41.66 & z: 33.33 & c: 8.33 \\
    y: 33.33 & t: 41.66 & u: 16.66 \\
    z: 33.33 & a: 33.33 & x: 8.33 \\
    \midrule
    \textbf{Average Accuracy: 30.44} & &  \\
    \bottomrule
    \end{tabu}%
  \label{tab:qwerty}%
\end{table}%

\subsection{Variations of the SH-NHHT and SH-HHT Attack Scenarios}
\label{subsec:variations}
\begin{figure}[t]
\vspace{-0.1in}
\centering
	\subfigure[]{
		\includegraphics[height= 1.25in]{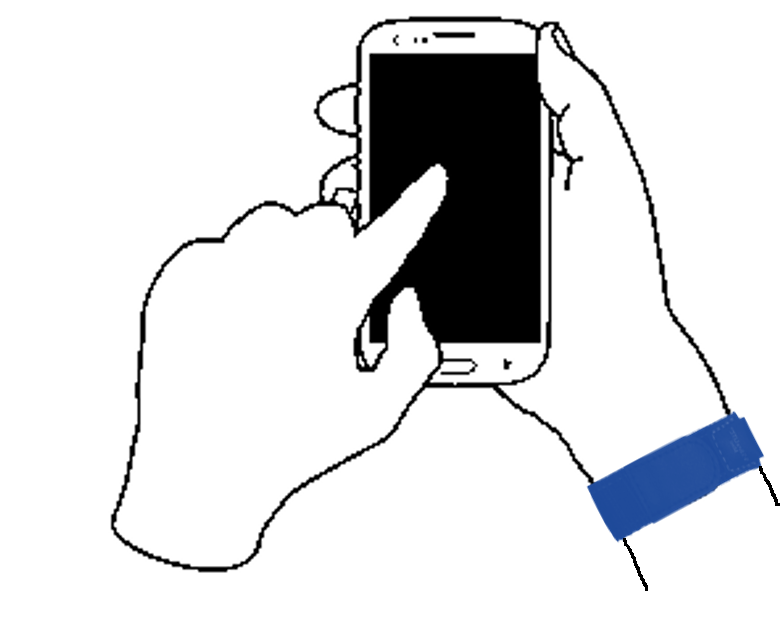}
		\label{fig:same-non-holding-mirror}
		}
	\hfill
	\subfigure[]{
		\includegraphics[height= 1.25in]{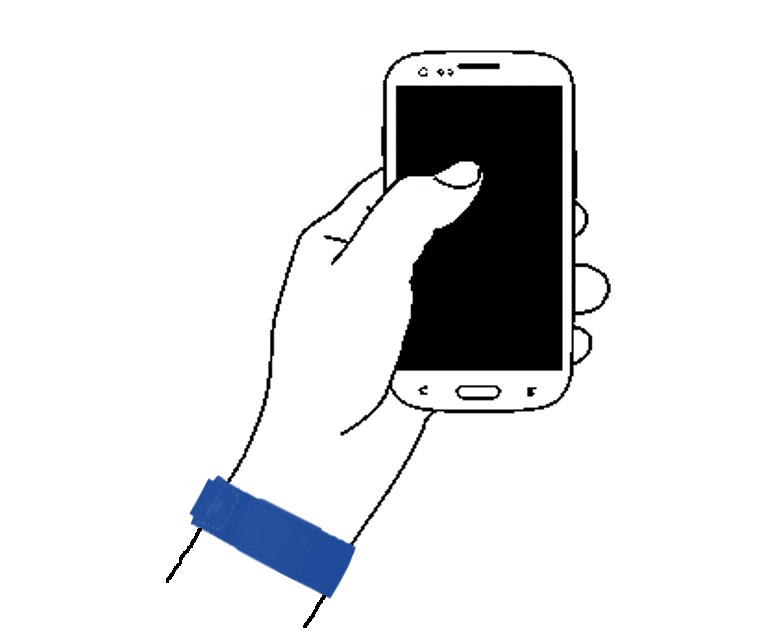}
		\label{fig:same-holding-mirror}
		}
\vspace{-0.15in}
\caption{Variations of typing scenarios in Figures \ref{fig:same-non-holding} and \ref{fig:same-holding}.}
\vspace{-0.0in}
\end{figure}

In addition to the SH-NHHT and SH-HHT scenarios presented in Figures \ref{fig:same-non-holding} and \ref{fig:same-holding}, there is an additional variation for each of these scenarios, as shown in Figures \ref{fig:same-non-holding-mirror} and \ref{fig:same-holding-mirror}. For SH-NHHT, the scenario \ref{fig:same-non-holding} assumes that the smartphone and smatwatch is on the left hand (and users type with the right hand). A variation of this SH-NHHT scenario is having the smartphone and smatwatch on the right hand and typing with the left hand (\ref{fig:same-non-holding-mirror}). A similar variation (\ref{fig:same-holding-mirror}) can also be envisioned for the SH-HHT scenario \ref{fig:same-holding}. We would like to analyze whether the performance of our proposed keystroke inference framework differs significantly for these variations. The same experiment as in Section \ref{ctc} was repeated by 12 new participants, each typing 100 randomly dictated numbers per variation. A two-tailed $t$-test on the \textit{One vs. Rest} classification accuracies for the SH-NHHT variations in Figures \ref{fig:same-non-holding} and \ref{fig:same-non-holding-mirror} returned the value of $t=-0.46$, $p=0.65$. For the SH-HHT variations \ref{fig:same-holding} and \ref{fig:same-holding-mirror}, it returned $t=0.61$, $p=0.54$. Both results are not significant at $p$<0.05, implying that our attack framework is not dependent on these variations. Therefore, an adversary can still use the same framework to carry out the inference attacks for these variations by simply retraining the classifiers. 


\section{Relative Transitions-Based Attack Framework}
\label{sec:expr2}

\begin{figure*}[]
\centering
\includegraphics[width=6.5in]{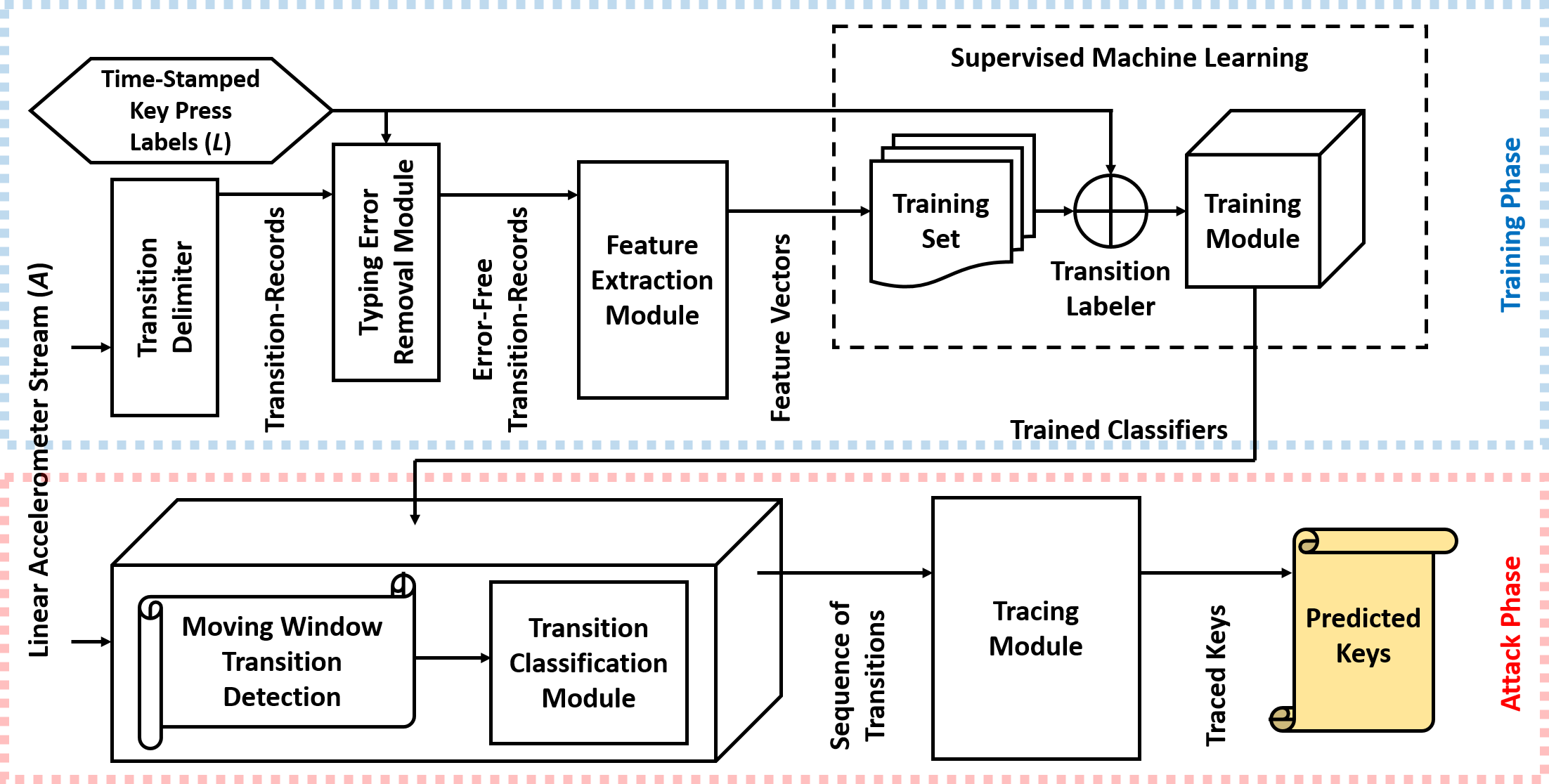}
\caption{\textbf{Overview of the relative transition-based attack framework for DH-NHHT typing scenario.}}
\label{fig:usingtransitions}
\end{figure*}

As discussed earlier, unlike the SH-NHHT and SH-HHT scenarios, key press events in the DH-HHT scenario (Figure \ref{fig:diff-non-holding}) cannot be uniquely and accurately detected on the smartwatch.
To overcome this problem in the DH-HHT scenario, we leverage on the observation that transitional movement between each pair of keys produces characteristically unique motions on the wrist, which can be accurately captured by the smartwatch. Accordingly, for the DH-HHT scenario, the keystroke inference framework (Figure \ref{fig:usingtransitions}) leverages on supervised machine learning to first classify transitional movements between consecutive key presses. Then, assuming a reasonable distribution of numbers typed, \textit{when multiple transitional directions in between a target sequence of key presses are traced on the key pad, we obtain a unique or highly reduced possibilities for the target sequence.}

\begin{figure}[]
\centering
\includegraphics[width= 3.4in]{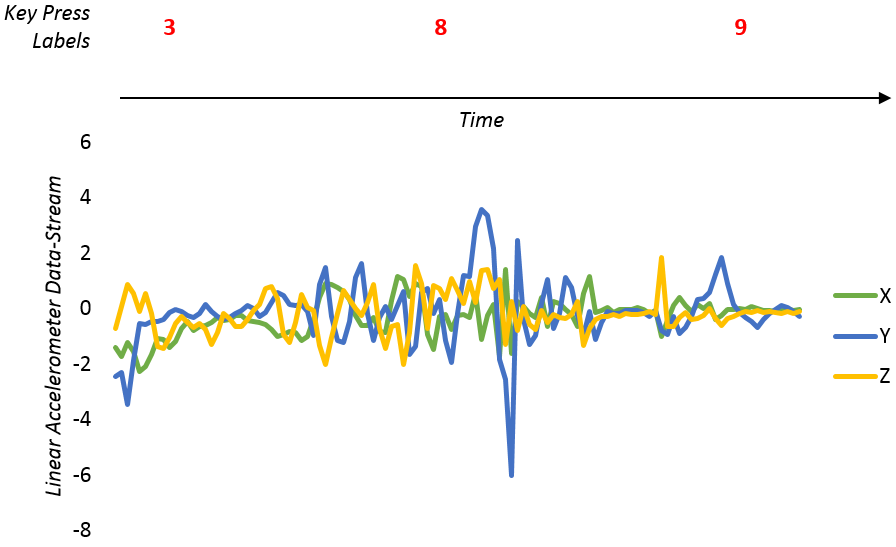}
\caption{\textbf{Time series of key press events in DH-NHHT, and their corresponding linear accelerometer readings. In DH-NHHT scenario, the wrist (along with the smartwatch) continues to move in between key press events. As a result, key press events cannot be identified or characterized based on spikes in energy level.}}
\vspace{-0.0in}
\label{transitionrecords}
\end{figure}

\textit{\textbf{Data Collection and Pre-Processing:}}
The same data collection application that was used for the SH-NHHT and SH-HHT scenarios is also used for the DH-NHHT scenario. Although the data collection process is exactly the same, the pre-processing operations are entirely different for DH-NHHT. Instead of detecting key press events, our goal here is to detect the type of wrist movement transition between every two consecutive key presses. As a result, we use the labeled stream of data to create labeled ``transition-records'' (Figure \ref{transitionrecords}) and use them as the training set. To create the training set, all linear accelerometer samples between two consecutive key press events are used as the transition-record.

\textit{\textbf{Transition Classification:}}
We classify transitions based on cardinal directions. The logic behind such a classification is that transitions in the same direction results in similar wrist movement. For example, wrist movement between numbers 4 and 1 would be similar to wrist movement between 6 and 3 (North), wrist movement between numbers 4 and 7 would be similar to wrist movement between 6 and 9 (South), and so on. One classifier is trained for each possible transitional direction, as listed in Table \ref{directions}: North (N), South (S), East (E), West (W), Northeast (NE), Northwest (NW), Southeast (SE), Southwest (SW) and Repeat (O). To achieve higher inference ability through tracing (explained later), the transition classifications must also be evenly populated. The number of possible transitions in each of the above nine categories follows a fairly even distribution, varying between 9 and 14. 

As an adversary will not have access to labels $L$, in the attack phase we use a variable-length moving window to check and determine the occurrences of transitions. The moving window is used to traverse (in steps of one sample) the linear accelerometer data $A$ in chronological order, and classify each window of linear accelerometer samples into one of the nine directions. The length of the window was varied from 10 samples (200 $msec$ at 50 Hz) to 100 samples (2 $sec$ at 50 Hz), to capture the variable length intervals possible between key presses. When ten or more consecutive windows were classified to be in the same direction, the classification result was recorded and the centroid was used as the key press time to form transition-records.

\begin{table}[b]
  \centering
  \caption{\textbf{Classification of all 100 possible numeric transitions.}}
  \vspace{0.05in}
    \begin{tabu} to \columnwidth {X[c]X[4]}
    \toprule
    N     & 4-1, 5-2, 6-3, 7-4, 8-5, 9-6, 0-8, 7-1, 8-2, 9-3, 0-5, 0-2, 0-1, 0-3 \\
    \midrule
    S     & 1-4, 2-5, 3-6, 4-7, 5-8, 6-9, 8-0, 1-7, 2-8, 3-9, 5-0, 2-0, 1-0, 3-0 \\
    \midrule
    E     & 1-2, 2-3, 4-5, 5-6, 7-8, 8-9, 1-3, 4-6, 7-9 \\
    \midrule
    W     & 2-1, 3-2, 5-4, 6-5, 8-7, 9-8, 3-1, 6-4, 9-7 \\
    \midrule
    NE     & 4-2, 5-3, 7-5, 8-6, 0-9, 4-3, 7-6, 7-2, 0-4, 8-3, 7-3 \\
    \midrule
    NW     & 5-1, 6-2, 8-4, 9-5, 0-7, 6-1, 9-4, 9-2, 0-6, 8-1, 9-1 \\
    \midrule
    SE     & 1-5, 2-6, 4-8, 5-9, 7-0, 1-6, 4-9, 1-8, 4-0, 2-9, 1-9 \\
    \midrule
    SW     & 2-4, 3-5, 5-7, 6-8, 9-0, 3-4, 6-7, 3-8, 6-0, 2-7, 3-7 \\
    \midrule
    O     & 1-1, 2-2, 3-3, 4-4, 5-5, 6-6, 7-7, 8-8, 9-9, 0-0 \\
    \bottomrule
    \end{tabu}
  \label{directions}
\end{table}

\textit{\textbf{Feature Extraction:}}
Contrary to the previous direct classification-based attacks, where each key press event was denoted in a fixed time period, transition periods between two key presses can vary widely depending on typing habit, keypad size, key pairs, etc. As a result, many of the time domain features used in SH-NHHT and SH-HHT scenarios cannot be applied for DH-NHHT. Thus, we rely mainly on frequency domain features, such as FFT of individual axis readings of the transition-record, their mean, correlation, spectral roll-off, spectral centroid, spectral flux and power spectral density estimates, to learn and classify transitions.

\textit{\textbf{Tracing and Recovery:}}
To infer a target sequence of key presses, the proposed framework tries to ``trace'' the transitions between key presses on the numeric keypad. Tracing eliminates all non-fitting key-pairs (the pair of keys that may have been pressed before and after a transition) for each transition of the target sequence, where the fitness of a key-pair is determined by the preceding and following transitions. In case tracing results in a uniquely identified key-pair for a transition, the keys pressed before and after that transition can be directly inferred. In other cases where tracing results in multiple possible key-pairs for a transition, the keys pressed before and after that transition can either be inferred by multiple trials or from the other adjoining key-pairs (only if the adjoining key-pairs are uniquely identified).

After the transitions are classified, tracing of keys can be performed using one of the following strategies:
\begin{itemize}[leftmargin=*]
\item
\textit{Forward Tracing:} The transitions are plotted on the keypad in the same order as they happened in time (Function $F\_Tracing()$ in Algorithm \ref{talgo}). In forward tracing, for a transition between candidate key pair $(p,q)$, if there does not exists a pair $(*,p)$ that satisfies the directional classification of the preceding transition, pair $(p,q)$ is eliminated from possible key pairs for that transition. The $F\_Tracing()$ function works from left to right on the test sequence.
\item
\textit{Backward Tracing:} The transitions are plotted on the keypad in the reserve order of how they actually happened in time (Function $B\_Tracing()$ in Algorithm \ref{talgo}). In backward tracing, for a transition between candidate key pair $(p,q)$, if there does not exists a pair $(q,*)$ that satisfies the directional classification of the following transition, pair $(p,q)$ is eliminated from possible key pairs for that transition. The $B\_Tracing()$ function works from right to left on the test sequence.
\item
\textit{Bidirectional Tracing:} Both forward and backward tracings are applied to reduce the possibilities for the target sequence (Function $BD\_Tracing()$ in Algorithm \ref{talgo}). 
\end{itemize}

We use bidirectional tracing in our evaluations because bidirectional tracing limits the propagation of any error that may be introduced by a transition misclassification.

\begin{algorithm}[h]
\caption{Tracing Algorithms}\label{talgo}
\begin{algorithmic}[]

\State $Transitions[N]$ ($N$ transitions in target sequence)
\State $Directions[]$ = $\{\emptyset\}$
\State $KeyPairs[]$ = $\{\emptyset\}$

\For {$i=1$ to $N$}
\State $Directions[i]$ = $Classify(Transitions[i])$
\State $KeyPairs[i]$ = $AllPossiblePairs(Directions[i])$
\EndFor
\State

\Function{F\_Tracing}{$KeyPairs[]$}
\For {$j=2$ to $N$}
\For{\textbf{each} pair $(p,q)$ in $KeyPairs[j]$}
\If {$\exists!$ a pair $(\ast,p)$ in $KeyPairs[j-1]$}
\State Remove $(p,q)$ from $KeyPairs[j]$
\EndIf
\EndFor
\EndFor\\
\Return $KeyPairs[]$
\EndFunction
\State

\Function{B\_Tracing}{$KeyPairs[]$}
\For {$k=N-1$ to $1$}
\For{\textbf{each} pair $(p,q)$ in $KeyPairs[k]$}
\If {$\exists!$ a pair $(q,\ast)$ in $KeyPairs[k+1]$}
\State Remove $(p,q)$ from $KeyPairs[k]$
\EndIf
\EndFor
\EndFor\\
\Return $KeyPairs[]$
\EndFunction
\State

\Function{BD\_Tracing}{$KeyPairs[]$}\\
\Return B\_Tracing(F\_Tracing($KeyPairs[]$))
\EndFunction

\end{algorithmic}
\end{algorithm}

\section{Evaluation of Relative Transition \\ Based Attack}
\label{evaluation2}

In this section, we present the findings from our evaluation of the relative transition based attack framework.

\subsection{Experimental Setup}
The same experimental setup and participants as in Section \ref{ref:exprsetup} were used for DH-NHHT. The only difference was that the smartwatch was worn on the right hand by the participants, and the smartphone was held in the left hand.

\subsection{Constructing and Testing the Framework}
\label{ctf}
\begin{figure}[]
\centering
\includegraphics[width = 3.45in]{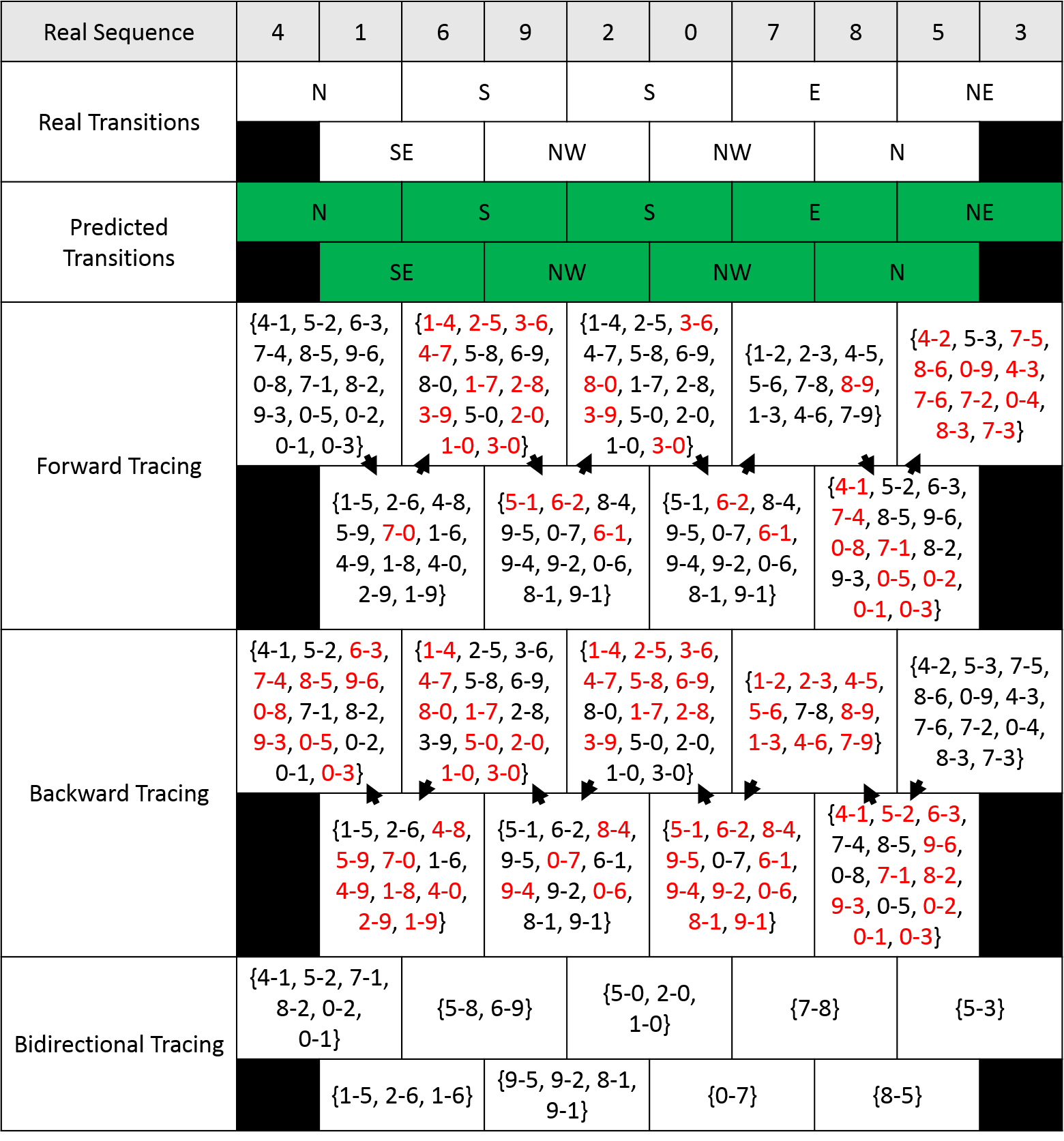}
\caption{\textbf{An example of how bidirectional tracing drastically reduces the possibilities of the key presses. First the forward tracing eliminates incompatible transitions (in red) in chronological order. Then the backward transition removes additional incompatible transitions in chronologically reverse order. In this example, we are able to uniquely identify the last 4  key-pairs using bidirectional tracing, which allows unambiguous inference of the last 5 key presses.}}
\label{tracing}
\vspace{-0.0in}
\end{figure}

We construct our transition classifiers based on training datasets of labeled transition-records generated by the same 12 participants who helped create the classifiers for the SH-NHHT and SH-HHT scenarios. The same audio stream of uniformly distributed random numbers between 0 to 9 guided the participants in typing 100 numbers. Out of the 1200 total numbers typed by all 12 participants, we use 960 numbers (having 948 transitions) for training  and rest for testing. We test the accuracy of the transition classifiers and tracing algorithms using two 10-digit long number sequences per participant (24 total test sequences, 240 total numbers, and 216 total transitions). We calculate the accuracy of the different tracing algorithms based on the number of correctly identified key presses in the traced number sequence. \textit{In order to infer a key, at least the preceding or following transitions should be uniquely identified.} For example, in the instance shown in Figure \ref{tracing}, the transition 9 to 2 and 2 to 0 both have other contending key-pairs (the incorrect transitions which are not removed by the tracing algorithm because they fit in the overall sequence of transitions). In such cases, it becomes impossible to determine the exact key (2 in this example) pressed in one trial. However, although the transition 2 to 0 have other contending key-pairs, the pressing of key 0 can be inferred with the help of the uniquely identified 0 to 7 transition, following the key press. In case both the preceding or following transitions are uniquely identified, the adversary can be more confident about the inference. One may also notice that the first and last number in a sequence are harder to infer, as there exists only one transition for each.

In our evaluation, the transition classifiers were able to correctly classify 191 transitions-records (88.42\% accuracy), while the remaining 25 incorrect or unclassified transitions introduced error in 17 of the test sequences. We also observe that an incorrect prediction is more likely to occur immediately after a previous incorrect prediction. One of the possible explanations behind such an observation is that the transition behavior varies depending on the preceding and following transitions. In terms of inference accuracy, 85 key presses out of the 240 test numbers were unambiguously identified using the bidirectional tracing (43.75\% accuracy). We relate the relatively low inference accuracy to three primary reasons: (a) incorrectly classified transitions introduce error in not one but two key presses, (b) unclassified transitions do not introduce error but there is no remedy to fill in the missing information, and (c) even a very small number of contending key-pairs makes it impossible to determine the exact key pressed.

\begin{figure}[]
\centering
\includegraphics[width = 3.2in]{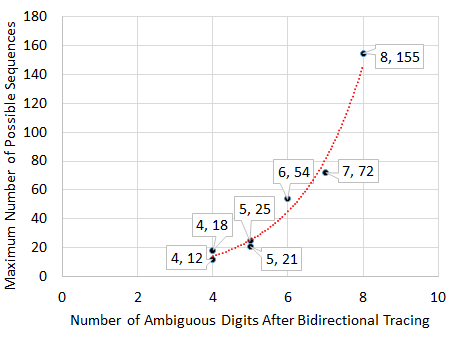}
\caption{\textbf{More ambiguously traced sequences require additional number of trials (in the worst case).}}
\label{moretrials}
\vspace{-0.0in}
\end{figure}

Because most of today's information systems acknowledge natural human mistakes and allows multiple trials to validate security tokens (pin, password, card number, etc.), adversaries can easily take advantage of it to try all possible number sequences derived from the output of bidirectional tracing. Accordingly, we evaluate the inference accuracy using multiple trials (solving from left to right), up to the maximum number of trials required to correctly infer the full number sequence. For example, in the instance shown in Figure \ref{tracing}, there can be 21 possible sequences derivable from the output of bidirectional tracing (listed in Table \ref{sequences}). Results of multiple trials are presented in Figure \ref{moretrials}, where we see that more ambiguous sequences require additional number of trials (in the worst case). We do not restrict the adversary to a certain number of attempts (which would be system dependent) because the actual sequence may or may not be tried in the limited number of attempts. Instead, we evaluate the worst case scenario, where the adversary has to try all possible sequences derived from the output of bidirectional tracing. Note that we evaluate this using only the 7 bidirectionally traced sequences for which all the predicted transitions are correct. 

\begin{table}[]
  \centering
  \caption{\textbf{The 21 possible number sequences that satisfy the bidirectional tracing obtained in Figure \ref{tracing}.}}
  \vspace{0.05in}
    \begin{tabu} to \columnwidth {X[c]X[c]X[c]}
    \toprule
    4-1-5-8-1-0-7-8-5-3 		& 7-1-5-8-1-0-7-8-5-3		& 0-1-5-8-1-0-7-8-5-3 \\
    4-1-6-9-5-0-7-8-5-3 		& 7-1-6-9-5-0-7-8-5-3		& 0-1-6-9-5-0-7-8-5-3 \\
    \textbf{4-1-6-9-2-0-7-8-5-3} 	& 7-1-6-9-2-0-7-8-5-3		& 0-1-6-9-2-0-7-8-5-3 \\
    4-1-6-9-1-0-7-8-5-3 		& 7-1-6-9-1-0-7-8-5-3		& 0-1-6-9-1-0-7-8-5-3 \\
    5-2-6-9-5-0-7-8-5-3 		& 0-2-6-9-5-0-7-8-5-3		& 8-2-6-9-5-0-7-8-5-3 \\
    5-2-6-9-2-0-7-8-5-3 		& 0-2-6-9-2-0-7-8-5-3		& 8-2-6-9-2-0-7-8-5-3 \\
    5-2-6-9-1-0-7-8-5-3 		& 0-2-6-9-1-0-7-8-5-3		& 8-2-6-9-1-0-7-8-5-3 \\
    \bottomrule
    \end{tabu}%
  \label{sequences}%
\end{table}%

\subsection{Combining Smartwatch and Smartphone Data}
As smartphones cannot capture transitional wrist movements of the typing hand, we cannot merge feature vectors like it was done in SH-NHHT and SH-HHT scenarios. As an alternative, we found a novel way to combine smartphone motion sensor data due to key taps, which was used in the classification-based attacks (as evaluated in Section \ref{smartphonecompare}), with the smartwatch transition-records obtained in DH-NHHT scenario. Based on the previous observation that classifying transition-records itself is highly accurate (88.42\% accuracy), we continue using the same attack framework. However, to overcome the limitations faced in the inference process (after tracing is completed), classified smartphone keystroke-records may be used to choose from the multiple candidate sequences obtained as the output of the tracing algorithm. We evaluate this attack using linear accelerometer readings of the smartphone, which were additionally collected during the DH-NHHT experiments of previous section. Keystroke-records and feature vectors are extracted from the smartphone data as it was done for SH-NHHT and SH-HHT. Elimination of contending key-pairs and filling up of undetected transitions with the help of classified smartphone keystroke-records (combined with reasonably accurate classification of keystroke-records), resulted in 82.50\% unambiguous inference of key presses. This is a substantial improvement in the inference accuracy, compared to the 43.75\% accuracy obtained earlier without the help of classified smartphone keystroke-records.

\subsection{A More Realistic Setting: Natural or Non-Controlled \\Typing}
\label{dhnatural}
Similar to SH-NHHT and SH-HHT experiments, the participants were being directed by an audio stream in the above DH-NHHT experiments, which may introduce a minor delay or disturbance in each key press. As a result, we conducted a similar natural or non-controlled typing experiment in the DH-NHHT scenario, where a completely new set of 12 participants were instructed to type their phone number followed by their residential zip code (15 numbers, 14 transitions). In this setting, out of a total of 168 transitions-records, 134 were classified correctly (79.76\% accuracy). The remaining incorrect or unclassified transitions introduced error in test sequences of 10 participants. Out of a total of 180 key presses, 69 were unambiguously identified using the bidirectional tracing, thus giving an accuracy of 38.33\%.

\section{Discussion}
\label{discussions}



\subsection{Limitations}

\noindent \textbf{Posture and Ambient Movement:} In practice, wrist movement patterns may change drastically based on the target user's body posture and orientation. In other words, the key press features while sitting may differ substantially from the key press features while laying down. One main limitation of our attack framework is that it is not robust against such different body postures and orientation. In order to overcome this, an attacker must train multiple classification models using data corresponding to different user postures and orientations, and then apply the appropriate one for the victim. This, if the attacker knows what was the victim's posture while typing. Similarly, if the target user is moving (for example: walking, running, sitting inside a car or train, etc.) while typing, keystroke events in the accelerometer/gyroscope data may get masked and our framework may not be able to correctly infer them. However, we must point out that this issue is not specific only to our attack framework, but other frameworks in the literature suffer from a similar drawback.

\noindent \textbf{Power Consumption:} Another limiting factor of our attack can be the power consumption rate on the smartwatch, due to the continuous recording of sensor data at a high frequency. For instance, the 300 mAh battery inside the Samsung Gear Live dropped from 100\% to 69\% in an hour, while recording linear accelerometer readings at 50 $Hz$. This limitation is less evident in case of smartphones due to their significantly higher battery capacity. To carry out a stealthy attack using the smartwatch, an attacker may have to either reduce the sensor sampling rate, or devise a mechanism to start the recording only when the potential victim is typing.

\noindent \textbf{Both Hand Typing:} We cover three major typing styles in this paper, while missing the case where a user holds the smartphone and types using both hands. In this scenario, the motion captured by the smartwatch will vary depending on which thumb is used to type a key, and which hand the smartwatch is worn on. The movement captured in this typing scenario will yield very different results and requires a new inference technique. We plan to work in this direction in the future.

\noindent \textbf{Threats to Validity:} Most of our experimental results were obtained from analysis of keystrokes typed in a relatively controlled setting, where participants were dictated on what to type. As a result, it is possible that those results may not be representative of how our attack framework may perform in more natural typing scenarios. However, we must point out that we do investigate the efficacy of our attack framework in several natural typing scenarios (in Sections \ref{realisticsetting} and \ref{dhnatural}), and the obtained results show that our inference framework has reasonable accuracy in these scenarios as well.

\subsection{Defenses}
Defending against side-channel attacks is a much debated topic \cite{cai2009defending}. Although modern mobile and wearable operating systems offer access control on some sensors, sensors such as accelerometer and gyroscope cannot be disengaged by the user. Moreover, most mobile applications do not require explicit permissions (either at install or run time) in order to access these sensors. A straightforward defense approach is to safeguard all sensors using system or user-defined access controls. However, such a static access control will become increasing complex to manage and will not protect against applications that gain legitimate access to these sensors. Reducing the frequency at which applications can sample data from these sensors is another potential defense mechanism. A system-level monitoring mechanism that tracks the context and frequency of sensor accesses, and appropriately flag unwanted accesses requested by applications, could also serve as a useful defense tool. 


\subsection{Enhancements}
\noindent \textbf{Random Walk Tracing:} This is a tracing algorithm we propose for use with very long number sequences typed in DH-NHHT scenario. In this tracing algorithm, a random subsequence of varying length is selected and bidirectional tracing is applied. The process is repeated several times such that every transition is covered multiple times, and each key press may end up having multiple candidate keys. Majority voting may be used to determine the final predicted keys (only if a key press has multiple candidate keys). This tracing algorithm will greatly minimize the propagation of any error that may be introduced by a transition misclassification.

\section{Conclusion}
\label{conclusion}
In this paper, we comprehensively investigated the feasibility of keystroke inference attacks on mobile numeric keypads by using smartwatch motion sensor data as an information side channel. We proposed two supervised learning-based frameworks to infer keystrokes from smartwatch motion data in three popular mobile holding and typing scenarios. We empirically evaluated the performance and efficacy of our proposed inference frameworks under various experimental settings (i.e., controlled versus natural typing), by using different types of smartwatch hardware, by using different types of motion sensors (i.e., accelerometer versus gyroscope) and by fusing motion data from multiple sources (i.e., smartphone and smartwatch). We also evaluated the performance of our attack framework on alphanumeric mobile keypads with a QWERTY layout. Results from our various experimental studies have shown that typing-induced motion data captured by smartwatch sensors can be employed as an effective side-channel to infer keystrokes on mobile keypads.


\section{Acknowledgments}
\label{ack}
We thank Ms. Kirsten Crager for helping us with keystroke data collection from human subject participants. This work has been supported by the Division of Computer and Network Systems (CNS) of the National Science Foundation (NSF) under award number 1523960.

\bibliographystyle{IEEEtran}
\bibliography{main}

\begin{IEEEbiography}[{\includegraphics[width=1in,height=1.25in,clip,keepaspectratio]{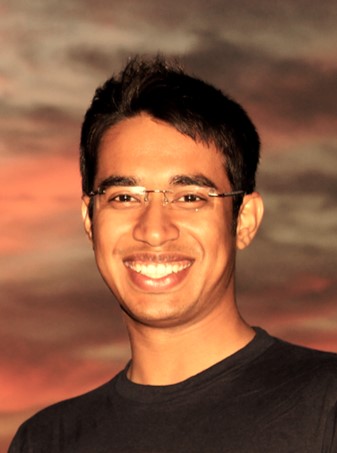}}]%
{Anindya Maiti}
received the M.S. degree in Electrical Engineering from Wichita State University, USA, in 2014, and the B.Tech. degree in Computer Science and Engineering from Vellore Institute of Technology, India, in 2012 . He is currently pursuing the Ph.D. degree in Electrical Engineering and Computer Science at Wichita State University, USA. His current research interests include cyber-physical system security and privacy.
\end{IEEEbiography}

\begin{IEEEbiography}[{\includegraphics[width=1in,height=1.25in,clip,keepaspectratio]{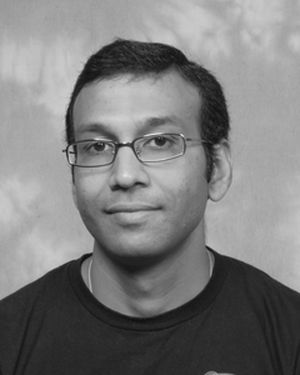}}]%
{Murtuza Jadliwala}
received his B.E. degree in Computer Engineering from Mumbai University, India, and his M.S. and Ph.D. degrees in Computer Science from the State University of New York at Buffalo, USA. He is currently an Assistant Professor in the Department of Electrical Engineering and Computer Science at the Wichita State University, USA. His current research focuses on overcoming security threats and privacy challenges in networked and cyber-physical systems.
\end{IEEEbiography}

\begin{IEEEbiography}[{\includegraphics[width=1in,height=1.25in,clip,keepaspectratio]{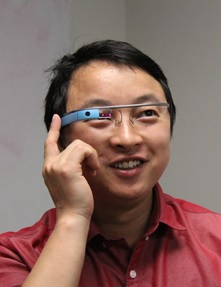}}]%
{Jibo He}
received his Ph.D. degree in Engineering Psychology from the University of Illinois, and Bachelor's degrees from Peking University. He is the director of the Human Automation Interaction Lab at Wichita State University. His research interests include human factors, driver distraction, aviation safety, eye movement, human-machine interaction, usability, and mobile devices.
\end{IEEEbiography}

\begin{IEEEbiography}[{\includegraphics[width=1in,height=1.25in,clip,keepaspectratio]{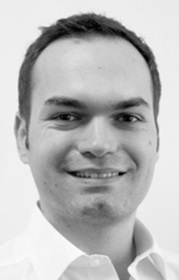}}]%
{Igor Bilogrevic}
is a Research Scientist at Google, which he joined in 2014. He earned his Ph.D. on the privacy of context-aware mobile networks from EPFL in 2014. From 2010 until 2012, he worked in collaboration with the Nokia Research Center on privacy in pervasive mobile networks, encompassing data privacy, social community privacy, location privacy and information-sharing. In 2013, he spent the summer at PARC (a Xerox Company), where he worked on topics related to private data analytics. His main research interests include applications of machine learning to privacy problems, private data analytics, contextual intelligence, and applied cryptography. He is co-inventor on several privacy-related patents.

\end{IEEEbiography}
\end{document}